\documentclass{article}
\usepackage{graphicx} 
\usepackage{graphicx,amsmath,amssymb,tikz,amsthm} 
\usepackage{arydshln}
\usepackage{mathtools}
\usepackage{enumerate}
\usepackage{bm}
\usepackage{cite}
\usepackage[linktocpage=true,
  colorlinks=true, 
  pdfborder={0 0 0},
  linkcolor=blue,
  citecolor=red,
  filecolor=yellow,
  urlcolor=blue,
  bookmarks,
  pdfauthor={},
]{hyperref}
\usetikzlibrary{quantikz2}
\usepackage{authblk}

\newtheorem{remark}{Remark}

\title{Real and Fourier space readout methods: Comparison of complexity and applications to CFD problems}

\author[1,2]{Xinchi Huang\footnote{Email: kkou@quemix.com; huangxc@g.ecc.u-tokyo.ac.jp}}
\author[1,2]{Hirofumi Nishi}
\author[1,2]{Yoshifumi Kawada}
\author[3]{Tomofumi Zushi}
\author[1,2,4]{Yu-ichiro Matsushita}
\affil[1]{Department of Physics, The University of Tokyo, Tokyo 113-0033, Japan}
\affil[2]{Quemix Inc., Taiyo Life Nihombashi Building, 2-11-2 Nihombashi Chuo-ku, Tokyo 103-0027, Japan}
\affil[3]{Sumitomo Rubber Industries, Ltd., 2-1-1 Tsutsui, Chuo, Kobe, Hyogo 651-0071, Japan}
\affil[4]{Quantum Materials and Applications Research Center, National Institutes for Quantum Science and Technology (QST), 2-12-1 Ookayama, Meguro-ku, Tokyo 152-8550, Japan}

\date{\today}

\begin{document}

\maketitle
\begin{abstract}
Quantum computing is a promising technology that accelerates the partial differential equations solver for practical problems. The reconstruction of solutions (i.e., the readout of quantum states) remains a crucial problem, although numerous efficient quantum algorithms have been proposed.  
In this paper, we propose and compare several efficient readout methods in the real and the Fourier space. The Fourier space readout (FSR) and the proposed approximate real space readout (ARSR) methods are currently the most efficient and practical ones for the purpose of reconstructing continuous real-valued functions. In contrast, the quantum amplitude estimation (QAE) based methods (especially in the Fourier space) are favorable for mid-term/far-term quantum devices. 
Besides, we apply the methods for benchmark solutions in computational fluid dynamics (CFD) and demonstrate great improvements compared to the conventional sampling method for large grid numbers. 
Equipped with efficient readout methods, we further show that a 2D Burgers' equation can be solved efficiently without using the expensive strategy of linearization. It suggests the potential quantum advantages for some practical applications on mid-term quantum devices.  
\end{abstract}

\section{Introduction}

Quantum computing is a revolutionary technology that has the potential to realize large matrix operations with possibly an exponentially small number of gate operations on quantum computers. 
Owing to the principles of quantum mechanics, unitary operations best fit quantum computing, and it was extensively discussed for chemistry problems \cite{Yuan2020, Lee.2023, Kassal.2008, Jones.2012, Ollitrault.2020, Childs.2022, Kosugi.2022, Kosugi.2023, Nishi.2023, Chan.2023, Brinet.2024, Huang.2024} where efficient Hamiltonian simulations are required. 
On the other hand, non-unitary operations can be realized by using ancillary qubits and post-selecting the ancillary qubits to be a desired state. In other words, a normalized non-unitary matrix can be regarded as a block of a larger unitary matrix. Thus, quantum computing has also attracted great attention in the field of engineering in which non-unitary operations are frequently used, and it is now regarded as a promising tool for computer-aided engineering (CAE) \cite{Kadowaki2025pre}. 

The core of the CAE simulations is the development of fast and accurate algorithms for solving the underlying partial differential equations (PDEs). The well-known Harrow-Hassidim-Lloyd (HHL) algorithm \cite{Harrow.2009} first revealed the potential exponential quantum speedup in the matrix size compared to the conventional classical methods (e.g., the sparse conjugate gradient method \cite{Shewchuk1994}). Since then, many advanced works, known as the quantum linear system algorithms (QLSAs), have improved the original HHL algorithm in the asymptotic behavior regarding the error bound and the condition number \cite{Ambainis2012, Berry2014, Berry.2017, Childs.2017, Childs.2020, Lin.2020, Childs.2021, An.2022, Costa.2022}. For the linear systems derived from the PDEs, the theoretical exponential speedup was confirmed by combining the quantum algorithms with the preconditioning techniques \cite{Clader.2013, Bagherimehrab.2023pre}. Besides, there are numerous works concerning the evolution equations (i.e., time-dependent PDEs) \cite{Kieferova.2019, Liu.2021, Kiani.2022, An.2022pre, An.2023, Fang.2023, Krovi2023, Jin.2023, Berry.2024, Sato.2024, Sato.2025, Jin.2024, Huang.2024pre}, which demonstrate polynomial/exponential speedup regarding the grid number. Although the quantum algorithms for general nonlinear PDEs were established in \cite{Krovi2023}, etc., the efficient and practical implementations on near-term or mid-term quantum devices have not yet been reported. Owing to this restriction, the quantum-classical hybrid algorithms, for example, the variational/non-variational differential equation solvers \cite{Bharadwaj.2023, Ye.2024, Chen.2024, Bharadwaj.2025, Song.2025, Zhuang.2025pre}, the quantum lattice Boltzmann method \cite{Sanavio.2024, Budinski2022, Li.2025, Itani.2024}, and some other approaches \cite{Gaitan2020, Kyriienko.2021, Meng.2023, Dewitte.2025}, are the majority for practical problems in computational fluid dynamics (CFD) (one of the typical CAE problems). We also refer to a review paper \cite{Malinverno.2025} and a fully quantum algorithm based on the quantum lattice gas algorithm \cite{Kocherla.2024}. 

The theoretical quantum speedup mentioned above was proved in the setting of preparing a quantum state corresponding to the PDE solution, while the readout of the quantum state (i.e., the reconstruction of the solution from the quantum state) was rarely touched because its cost was believed to be linear regarding the nonzero components of the state, which could eliminate the achieved speedup, especially in the quantum-classical hybrid algorithms in which frequent exchanges of quantum and classical information are required. 
Recently, the end-to-end exponential speedup, including the readout part, was confirmed \cite{Huang.2025pre, Zhuang.2025pre} using the readout in the Fourier space. Thanks to the fact that the solutions for many practical CAE problems are, to some extent, continuous and differentiable, the Fourier coefficients become localized, which is the key reason why the readout can be efficient. 
Beyond the readout in the Fourier space, there are some other basis function expansion (feature extraction) methods \cite{Nishi.2025pre, Miyamoto.2023} based on the optimization of weights on classical computers using the calculated overlaps on quantum computers. Roughly speaking, such optimization-based methods have the quantum complexity $O(M^2 N_\text{iter})$, while the sampling-based method (e.g., FSR in \cite{Huang.2025pre}) has the complexity $O(M)$ where $M$ is the number of effective weights in the given bases and $N_{\text{iter}}$ is the iteration number in the optimization. This implies that the optimization-based method is efficient for small $M$\footnote{In general cases, the pre-factor of the optimization-based method is usually smaller than that of the sampling-based method because we need only the state preparations of the basis vectors for the former one, while we need to implement the transformation matrix between the bases for the latter one. It does not change much for the Fourier basis since both the vector and the matrix are implemented by the quantum Fourier transforms.}. 
Besides, we refer to the quantum machine learning (QML) approaches, e.g., \cite{Williams.2024pre} and the references therein, for a CFD classification problem, where the solutions themselves are not essential.

In this paper, we investigate the practical readout methods, especially for the use case in CFD simulations. 
Our first contribution is the proposal and comparison of several readout methods in the real and the Fourier spaces. The theoretical analysis on the complexity in Table \ref{sec3:tab1} implies that a specific real space readout method eliminates the dependence on the grid number. Besides, the efficiency of the readout methods in the Fourier space highly relies on the regularity/smoothness and periodicity of solutions, which means the Fourier space readout methods are remarkable for smooth periodic functions.  
The second contribution is the case study for the benchmark solutions in the CFD simulations. Although still $10^5\sim 10^7$ shots are required to obtain relatively good solutions, the proposed grid-number-independent methods greatly outperform the conventional method, and hence the quantum advantage survives (especially in computational time, see Remark \ref{sec3:rem3}). 
Finally, in terms of the efficient readout methods, we show that a two-dimensional (2D) Burgers' equation (as an example of nonlinear evolution equations) could be efficiently solved by a time marching method using the time stepwise readout (TSR) strategy. Compared to the Carleman linearization strategy used in \cite{Krovi2023}, the circuit depth and the number of qubits are largely reduced. This implies that the TSR strategy is the more practical approach for the near-term and mid-term quantum devices. 

The paper is organized as follows. 
In Sect.~\ref{sec:2}, we propose several efficient readout methods: an approximate real space readout method ARSR, a modified Fourier space readout method FSR, and a combined method using quantum amplitude estimation (QAE) in the Fourier space FSQAE (see Section \ref{subsec:2-3}). 
Next, we analyze and summarize the theoretical complexities for the methods in Sect.~\ref{subsec:3-1} whose orders are verified by numerical examples in Sect.~\ref{subsec:3-2}. 
Besides, the feature of each method is provided in Sect.~\ref{subsec:3-3} so that we can choose a suitable method according to the given problem. 
Section \ref{sec:4} is devoted to the applications to the CFD simulations. The visualizations of the CFD solutions in two benchmark problems are addressed in Sect.~\ref{subsec:4-1}. Moreover, we propose a TSR strategy combining the efficient readout methods with an approximate probabilistic imaginary-time evolution (PITE) algorithm \cite{Huang.2024pre} to solve a 2D Burgers' equation in Sect.~\ref{subsec:4-2}.  
Section \ref{sec:5} summarizes the results and mentions several future topics to extend this work further. Some theoretical details, as well as numerical details, are provided in Appendices \ref{sec:appA}--\ref{sec:appF}.  


\section{Methods}
\label{sec:2}

In this section, we list several real and Fourier space readout methods targeted at recovering the real-valued functions that are encoded in given quantum states. 
Here, we address the methods without solving optimization problems to avoid the costs of the optimization loops. 

Let $d\in \{1,2,\ldots\}$ be the spatial dimension, $n_\ell\in \mathbb{N}$ be the number of qubits for each dimension. Set the total number of qubits to be $n_{\text{tot}}=\sum_{\ell=1}^d n_\ell$ and the grid numbers to be $N_\ell = 2^{n_\ell}$, $N=\prod_{\ell=1}^d N_\ell = 2^{n_{\text{tot}}}$. 
The input quantum state is given as follows:
$$
\ket{\psi}_{n_{\text{tot}}} = \sum_{j=0}^{N-1} \psi_j \ket{j}_{n_{\text{tot}}}, \quad \psi_j\in \mathbb{R},\ j=0,\ldots,N-1.
$$
Our aim is to obtain the approximations of the coefficients $\{\psi_j\}$ at all the grid points $j\in \{0,\ldots,N-1\}$ or some target grid points $j\in \mathcal{J}\subset \{0,\ldots,N-1\}$. More precisely, we consider an underlying (continuous) real-valued function $f: [0,L_1] \times \cdots \times [0,L_d]\rightarrow \mathbb{R}$ where $L_\ell>0$ is the length of the rectangular domain in each dimension. Then, the coefficients satisfy 
$$
\psi_j = \frac{f(\mathbf{x}_j)}{A_N}, \quad \mathbf{x}_j = (j_1 L_1/N_1,\ldots,j_d L_d/N_d),\ j_\ell=0,\ldots,N_\ell-1,
$$
where $A_N := \left(\sum_{j=0}^{N-1}|f(\mathbf{x}_j)|^2\right)^{1/2}$ is the normalized factor. Here, we use the correspondence: 
$$
j=\sum_{\ell=1}^{d} j_\ell \left(\prod_{\ell^\prime=1}^{\ell-1} N_{\ell^\prime}\right)
$$ 
between an integer $j$ and a vector $\mathbf{j} = (j_1,\ldots,j_d)$.

\subsection{Revisits: Real space readout (RSR) and Fourier space readout (FSR)}
\label{subsec:2-1}

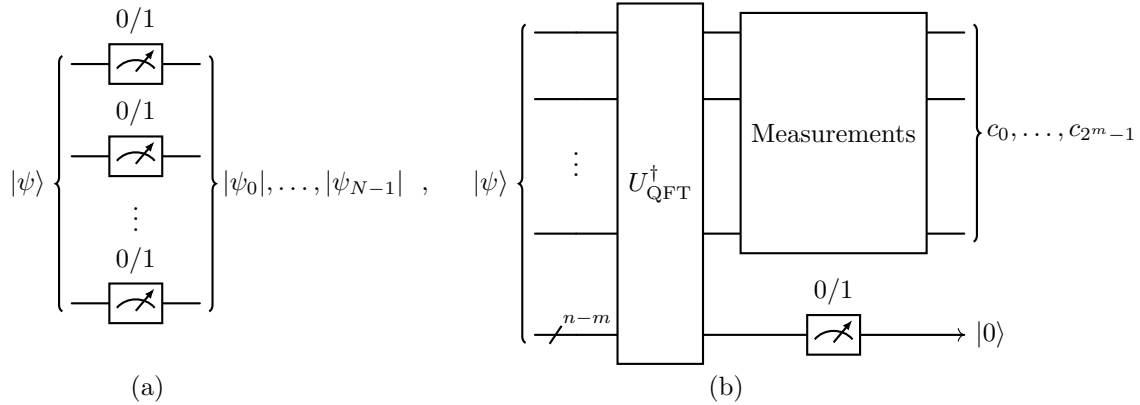
\begin{figure}
\centering
\begin{quantikz}
\lstick[4]{\ket{\psi}} & \meter{0/1} & \rstick[4]{$|\psi_{0}|,\ldots,|\psi_{N-1}|$} \\[0.2cm]
 & \meter{0/1} & \\[-0.3cm]
\setwiretype{n} & \vdots & \\[0.2cm]
 & \meter{0/1} & 
\end{quantikz}
, \quad
\begin{quantikz}
\lstick[5]{\ket{\psi}} &  & \gate[5]{U_{\text{QFT}}^\dag} & \gate[4]{\text{Measurements}} & \rstick[4]{$c_{0},\ldots,c_{2^m-1}$} \\[-0cm]
 &  &  &  &  \\[-0cm]
\setwiretype{n} & \vdots & \vdots &  & \\
 &  &  &  &  \\[0.2cm]
 & \qwbundle{n-m} &  & \meter{0/1}\arrow[r] &
\rstick{\ket{0}} 
\end{quantikz}
\\
(a) \hspace{7cm} (b) \hspace{3.5cm}
\caption{(a) Quantum circuit for the real space readout. (b) Quantum circuit for the Fourier space readout with an integer parameter $m\le n$. The first $2^m$ Fourier coefficients $c_0,\ldots,c_{2^m-1}$ can be obtained by post-selecting the most significant $n-m$ qubits to be in $\ket{0}_{n-m}$ state.}
\label{sec2:fig1}
\end{figure}

A direct way is to execute $Z$-basis measurements for all the qubits, and then, by the repetitions of measurements, we can construct a histogram to obtain the absolute square of each amplitude, that is, $|\psi_j|^2$ for $j=0,\ldots,N-1$, if only the absolute values are of interest. Considering the readout of a real-valued function at grid points, we call this basic method the real space readout (RSR) in this paper, see Fig.~\ref{sec2:fig1}(a).  

As mentioned in our previous work \cite{Huang.2025pre}, the RSR method is not good for large $N$ because the number of repetitions needs to be $O(N)$ in general. To avoid the dependence on the grid number $N$, we proposed a Fourier space readout (FSR) method. A similar idea is also used in a recent paper \cite{Zhuang.2025pre} for the CFD simulations. 
The key idea is to use a quantum Fourier transform (QFT), and by introducing an approximation parameter $m<n$, one can obtain the first $2^m$ Fourier coefficients, which is demonstrated in Fig.~\ref{sec2:fig1}(b). According to the Fourier series expansion of the function, we can reconstruct the function approximately by 
\begin{equation}
\label{sec2:eq-cfr}
f(\mathbf{x}) \approx C_N\sum_{k_1=-(M_1-1)}^{M_1-1}\cdots \sum_{k_d=-(M_d-1)}^{M_d-1} c_{\mathbf{k}} \prod_{\ell=1}^d \mathrm{exp}\left(\mathrm{i}\frac{2\pi}{L_\ell} k_\ell x_\ell\right), \quad \mathbf{x}\in [0,L_1] \times \cdots \times [0,L_d],
\end{equation}
where $M_\ell=2^{m_\ell}$, and $m_\ell$, $\ell=1,\ldots,d$ is the approximation parameter for each dimension. Here, the constant $C_N=A_N/\sqrt{N}$. 
The required number of shots is independent of $N$, and the FSR method greatly outperforms the RSR method in the cases of continuous functions owing to the rapid decay of the Fourier coefficients \cite{Huang.2025pre}. 

Since obtaining a series of complex numbers is not easy, we used some extension operators to make the Fourier coefficients of the extended function real \cite{Huang.2025pre}, so that the fundamental Z-basis measurements are enough to obtain the coefficients. Here, we slightly modify the quantum circuits and provide an alternative implementation of the FSR method without using the extension operator (see Appendix \ref{sec:appA}). Although the gate complexity has the same order as the original FSR method in \cite{Huang.2025pre}, we reduce the number of ancillary qubits from $d+1$ to $2$. In addition, the modified FSR method helps to avoid the possible artificial ``singularity" introduced by the extension, and hence, has better performance than the original one for some functions (especially the periodic functions with non-zero spatial derivatives on the boundary). 
In this paper, we use the modified FSR method for the numerical examples and simply denote it the FSR method if there is no confusion. 

For many practical applications, the readout in Fourier space is usually more flexible and efficient than that in real space because we obtain the dominant ``features" of the underlying function instead of only the values at the grid points. In the scenarios where we intend to calculate the values in some target domains or some weighted integral of the function, we do not need to know all the values at the computational grid points. Then, the total cost is completely independent of $N$. 

\subsection{Approximate RSR method (ARSR)}
\label{subsec:2-2}

\begin{figure}
\centering
\begin{quantikz}
\lstick[5]{\ket{\psi}} & \qwbundle{\!\!n_1-m_1} &  &  \\[0.2cm]
 & \qwbundle{m_1} & \meter{0/1} & \\[-0.3cm]
\setwiretype{n} & \vdots & \vdots & \\[0.1cm]
 & \qwbundle{\!\!n_d-m_d} &  & \\[0.2cm]
 & \qwbundle{m_d} & \meter{0/1} & 
\end{quantikz}
\caption{Quantum circuit for the RSR method with approximation parameters $m_1,\ldots,m_d$. Z-basis measurements are executed for the dominant qubits in each spatial dimension. }
\label{sec2:fig2}
\end{figure}
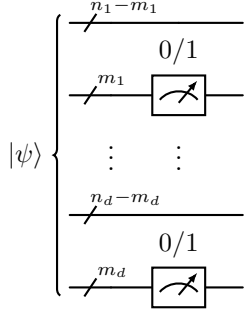

If the function is known to be positive in advance, then we can eliminate the $N$-dependence of the RSR method at the cost of some additional approximation error. The idea is to measure only a limited number of qubits to obtain approximate values at coarser grid points and then evaluate the values at the target points by interpolation (e.g., spline interpolation). 
More precisely, we propose the quantum circuit in Fig.~\ref{sec2:fig2}. 
By measuring only the dominant $m_1+\cdots+m_d$ qubits, the estimated amplitudes do not depend on $N$. Here, $m_1,\ldots,m_d$ are some suitably chosen approximation parameters smaller than the corresponding grid parameters $n_1,\ldots,n_d$. 
It is readily seen that the result of the quantum circuit in Fig.~\ref{sec2:fig2} gives a root mean square (RMS) approximation of the values at $2^{m_1+\cdots+m_d}$ coarse grid points. Therefore, the result of this approximation method is equivalent to the result of the RSR method after a specific post-processing of taking the RMS values in the coarse grid domains. Provided that the function is absolutely continuous, the theoretical analysis of the RMS approximation error shows that the approximation parameters $m_1,\ldots,m_d$ depend on the error bound, but are independent of $N$. 
We emphasize that unsuitable post-processing (e.g., taking the mean value) of the RSR result could amplify the statistical error. Several post-processing methods are compared in Appendix \ref{subsec:C2-1}, and we conclude that only the proposed RMS post-processing is efficient for large grid numbers. Such a specific RMS post-processing is good as it relieves the effect of the statistical error. 
For convenience, we call the approach using the quantum circuit in Fig.~\ref{sec2:fig2} (or equivalently applying the RMS post-processing after using the quantum circuit in Fig.~\ref{sec2:fig1}(a)) the approximate real space readout (ARSR) method in this paper.  
Different from the FSR method, the choice of the approximation parameters $m_1,\ldots,m_d$ for the ARSR is not automatically determined by the number of shots, although we can estimate its order regarding the grid number and the error bound. In practice, we gradually increase the parameters and check whether the change of the $l^2$ norm is convergent or not to determine the parameters (see Remark \ref{sec2:rem1}). 

\subsection{QAE-based methods}
\label{subsec:2-3}

It is well-known that the quantum amplitude estimation (QAE) based on the quantum amplitude amplification (QAA) achieves a quadratic speedup regarding the error bound, compared to the direct sampling methods. We combine the QAE algorithm with either the real space readout or the Fourier space readout to relieve the cost regarding the error bound. 

First, we apply the well-performed RQAE (real QAE) algorithm proposed in \cite{Manzano.2023} to estimate the amplitudes at the target grid points $\mathbf{x}_j$, $j\in \mathcal{J}$. Assume that we are given an oracle to access the input quantum state:
$$
U_{\psi} \ket{0}_{n_{\text{tot}}}:= \ket{\psi}_{n_{\text{tot}}}.
$$
For each $\mathbf{x}_j$, we apply the inverse of the quantum modular adder $U_{\mathrm{MADD}}^\dag[j]$ (e.g., \cite{Li.2020} with less gate count, \cite{Yuan.2023} with less ancillary qubits) to the oracle $U_\psi$ and obtain a shifted operation $U_{\psi_j}$:
$$
U_{\psi_j} \ket{0}_{n_{\text{tot}}} := U_{\mathrm{MADD}}^\dag[j] U_{\psi} \ket{0}_{n_{\text{tot}}} = \sum_{k=0}^{N-1} \psi_k \ket{k-j}_{n_\text{tot}} = \psi_j \ket{0}_{n_\text{tot}} + \sum_{k=1}^{N-1} \psi_{(k+j)\, \mathrm{mod}\, N} \ket{k}_{n_\text{tot}}.
$$
Regarding $\ket{0}_{n_\text{tot}}$ as a good state, and noting that the rest part is perpendicular to the zero state, we can apply the QAE algorithm in \cite{Manzano.2023} to estimate the amplitude $\psi_j$ for each $j$. 
The quantum circuit is provided in Appendix \ref{sec:appB}. In this paper, we denote such a method by RSQAE for convenience.


Next, similar to the QAE-based method in the real space, we can also consider the QAE-based method in the Fourier space. For arbitrary $k\in \{0,\ldots,N-1\}$, we introduce the shifted operation based on the given oracle as follows:
$$
U_{c_k} \ket{0}_{n_{\text{tot}}} := U_{\mathrm{MADD}}^\dag[k] \left(\bigotimes_{\ell=1}^d U_{\text{QFT}}^\dag\right) U_{\psi}  \ket{0}_{n_{\text{tot}}} = c_k \ket{0}_{n_\text{tot}} + \sum_{j=1}^{N-1} c_{(j+k)\, \mathrm{mod}\, N} \ket{j}_{n_\text{tot}}. 
$$
Using the extension operator in the FSR method in \cite{Huang.2025pre}, we have real-valued Fourier coefficients. Hence, we can again employ the RQAE algorithm in \cite{Manzano.2023} to obtain the Fourier coefficients for each $k$ that we need to calculate. For completeness, we provide the detailed quantum circuits in Appendix \ref{sec:appB}. In this paper, we denote such a QAE-based readout method in the Fourier space with the extension operator by FSQAE. We mention that one can also combine the idea of the modified FSR method in Appendix \ref{sec:appA} with the QAE algorithm. If the extension does not change the regularity of the function, then the QAE-based readout method using the idea of the modified FSR would be slightly worse than the FSQAE because we need twice the QAE algorithms to obtain both the real part and the imaginary part of the Fourier coefficients. 
\begin{remark}[Determinations of the approximation parameters]
\label{sec2:rem1}
As described above, we introduced the approximation parameters $m_\ell$, $\ell=1,\ldots,d$ for the ARSR, FSR, and FSQAE methods. 
The parameters for the FSR method could be adaptively determined according to the number of shots \cite{Huang.2025pre}. 
For the ARSR method, we start from small $m_1=\cdots=m_d=1$ and calculate the approximate function $f_{(1)}$. For $1\le k\le \min_\ell{n_\ell}-1$, we let $m_1=\cdots=m_d=k+1$ and calculate the approximate function $f_{(k+1)}$. In addition, we calculate the $l^2$ norm between the two approximate functions: $\mathrm{err}_k :=\|f_{(k+1)}-f_{(k)}\|$ to check whether the approximate function is convergent. If $\mathrm{err}_k > \mathrm{err}_{k-1}$, then we regard $f_{(k)}$ as the proper approximate function and $m_1=\cdots=m_d=k$ as the approximation parameters. Otherwise, we let $k=k+1$ and calculate the next approximate function. 
For the FSQAE method, it is unclear how to determine the parameters. In this paper, we take them empirically, e.g., $m_1=\cdots=m_d=6$. 
\end{remark}

\section{Complexity}
\label{sec:3}

We discuss the theoretical overheads of the complexity for the methods in Sect.~\ref{sec:2}. Besides, we numerically check the performance of the methods for several simple functions.  

\subsection{Theoretical bounds}
\label{subsec:3-1}

In the following context, we give a brief explanation of the complexity for each method, and the theoretical details are provided in Appendix \ref{sec:appC}.

First, we consider the sampling-based methods whose complexity is proportional to the number of shots. 
For the RSR method, the required number of shots is $O(N/\varepsilon^2)$ where $\varepsilon>0$ is the error bound. Here, the dependence on the grid number $N$ comes from the fact that $\psi_j$ is proportional to $1/\sqrt{N}$. 
For the FSR method, the Fourier coefficients are uniform regarding $N$, which yields an $N$-independent overhead on the number of shots \cite{Huang.2025pre}. Roughly speaking, the number of shots is $O(M/\varepsilon^2)$ where $M:=\prod_{\ell=1}^d M_\ell=O(1/\varepsilon^{ds})$ is the number of dominant Fourier coefficients. Here, we assume the decay of the Fourier coefficients $c_{\mathbf{k}} = O(k_1^{-p}\cdots k_d^{-p})$ for some integer $p\ge 1$ and use the regularity parameter $s:=2/(2p-1)\in (0, 2]$ to indicate the smoothness of the function. For a continuous piecewise $W^{2,1}$ (i.e., up to the second-order derivatives are integrable) function, we have $p\ge 2$, and hence, $s\le 2/3<1$. By a careful estimation, the required number of shots is improved to $\tilde{O}(1/\varepsilon^{2+s})$, whose power of $1/\varepsilon$ has no explicit dependence on the dimension $d$, see Remark \ref{appC:rem6} in Appendix \ref{sec:appC}. Here and henceforth, by the notation $\tilde{O}(1/\varepsilon^{2+s})\equiv O((1/\varepsilon^{2+s})\mathrm{polylog}(1/\varepsilon))$, we omit the minor dependence on the key parameters (i.e., error bound and grid number) for simplicity. 
For the ARSR method, we also introduced some approximation parameters $m_1,\ldots,m_d$. Letting $M_\ell=2^{m_\ell}$, $\ell=1,\ldots,d$, we define $M:=\prod_{\ell=1}^d M_\ell$. The number of shots is $O(M/\varepsilon^2)$ using a similar discussion to the RSR method. Considering the RMS approximation error, we have $M=O(1/\varepsilon^{d/2})$ in the case that $f$ is smooth enough (e.g., $f\in C^2$), and $\partial_\mathbf{x}f(\mathbf{x_0})=0$ for any $\mathbf{x_0}$ satisfying $f(\mathbf{x_0})=0$. 

On the other hand, the complexity of the QAE-based methods should be regarded as the number of calls to the oracle $U_\psi$. Employing the well-performed QAE algorithm in \cite{Manzano.2023}, we need $O(\log(1/\epsilon_0))$ iterations to obtain each $\psi_j$ or $c_k$ up to an error bound $\epsilon_0$ using $\tilde{O}(1/\epsilon_0)\equiv O((1/\epsilon_0)\mathrm{polylog(1/\epsilon_0)})$ queries to the oracle. 
For the RSQAE method, we have $\epsilon_0=O(\varepsilon/\sqrt{N})$, which gives the complexity $\tilde{O}(J \sqrt{N}/\varepsilon)$ where $J\le N$ is the number of target grid points. 
For the FSQAE method, we have $M=O(1/\varepsilon^{ds})$ considering the approximation error (Fourier truncation error). By balancing the QAE and the approximation errors, we obtain $\epsilon_0 = O(\varepsilon^{1+ds/2})$. This yields the complexity $O(1/\varepsilon^{1+3ds/2})$.

We summarize the above results in Table \ref{sec3:tab1}. Here, $J$ is the number of target grid points, and $\mathcal{C}_\psi$ denotes the circuit depth of the oracle $U_\psi$ for preparing the input quantum state. To achieve a quantum speedup in $N$, it is expected that $\mathcal{C}_\psi = O(\mathrm{polylog} N)$.  
\begin{table}[htb]
\centering
\caption{Quantum and classical complexities for the real space and Fourier space methods. }
\label{sec3:tab1}
\scalebox{0.85}[0.85]{
\begin{tabular}{l|ccc}
\hline
Method & Maximal circuit depth & Quantum complexity & Classical complexity \\
\hline
RSR$^{[1]}$ & $O(\mathcal{C}_\psi)$ & $O\left(\mathcal{C}_\psi N (1/\varepsilon)^{2}\right)$ & $O\left(N (1/\varepsilon)^{2}\right)$ \\
ARSR$^{[1]}$ & $O(\mathcal{C}_\psi)$ & $O\left(\mathcal{C}_\psi (1/\varepsilon)^{2+d/2}\right)$ & $O\left((1/\varepsilon)^{2+d/2}\right) + O\left(J(1/\varepsilon)^{d/2}\right)$ \\
FSR & $O(\mathcal{C}_\psi+\mathrm{polylog} N)$ & $\tilde{O}\left((\mathcal{C}_\psi+\mathrm{polylog} N) (1/\varepsilon)^{2+s}\right)$ & $\tilde{O}\left((1/\varepsilon)^{2+s}\right) + O\left(J(1/\varepsilon)^{ds}\right)$ \\
RSQAE & $O(\mathcal{C}_\psi \sqrt{N}/\varepsilon)$ & $\tilde{O}\left(\mathcal{C}_\psi J\sqrt{N}(1/\varepsilon)\right)$ & $O\left(J\mathrm{polylog}\left(\sqrt{N}/\varepsilon\right)\right)$ \\
FSQAE & $O\left((\mathcal{C}_\psi+\mathrm{polylog} N)/\varepsilon^{1+ds/2}\right)$ & $\tilde{O}\left((\mathcal{C}_\psi+\mathrm{polylog} N) (1/\varepsilon)^{1+3ds/2}\right)$ & $\tilde{O}\left((1/\varepsilon)^{ds}\right) + O\left(J(1/\varepsilon)^{ds}\right)$ \\
\hline
\end{tabular}
}
\begin{flushleft}
\footnotesize [1] The complexity is obtained under the assumption that the function has a fixed sign. Moreover, we assume that $f$ is smooth enough and satisfies $\partial_\mathbf{x}f(\mathbf{x_0})=0$ if $f(\mathbf{x_0})=0$ for the ARSR method. Without such assumptions, the order of the ARSR method can be $O((1/\varepsilon)^{2+d})$ in general. 

\end{flushleft}
\end{table}
In the previous work \cite{Huang.2025pre}, we discussed the complexities of the RSR and the FSR methods in a one-dimensional case. By Table \ref{sec3:tab1}, we show how the spatial dimension $d$ would influence the power of $1/\varepsilon$ where we omit the minor $d$-dependence in the pre-factors. In addition, the complexities of the other proposals mentioned in Sect.~\ref{sec:2} are given for comparison. 

Paying attention to the second column, we find that the RSR and the ARSR methods have the least circuit depth, while the quantum circuits for the FSR method are slightly deeper. On the other hand, the QAE-based methods have much larger depths, depending on the error bound or the grid number, because of the use of the QAA operators. Noting that deeper quantum circuits require real quantum devices with smaller quantum noises, the RSR, the ARSR, and the FSR methods are expected to work on the real devices before 2030 according to the roadmaps of the quantum hardwares \cite{IBM, Quantinuum}. Mid-term/far-term devices are required for the QAE-based methods. 

Referring to the third column, the ARSR and the readout methods in the Fourier space are more efficient regarding the grid number $N$. If the function is sufficiently smooth such that $s<d/2$, then the FSR method outperforms the ARSR method. Besides, the FSQAE method has smaller complexity than the FSR method if $s<2/(3d-2)$. 
Moreover, we can translate the grid number into the error bound: $N=O(1/\varepsilon^d)$ if high precision is required. In such a situation, the ARSR method outperforms the RSR method, the Fourier space readout methods are better than the ARSR method provided that $s<d/2$, and FSQAE is better than the FSR if $s<2/(3d-2)$. Note that the quantum complexity of the RSQAE method depends on the number of target points. It has small complexity if $J=O(1)$ (i.e., limited target points independent of $N$ and $\varepsilon$), while it has the largest complexity if $J=N$.

The classical complexity in the fourth column includes both the computational cost in making the histogram of the quantum measurements (the first term) and the computational cost of reconstructing the value of the function at the target points (the second term if applicable). Since the reconstruction can be done in parallel for each target point, the computational time of the ARSR, the FSR, and the FSQAE methods is independent of $N$ even when $J=N$. In fact, if we take the execution time of a classical operation and a quantum operation on current devices into account, such calculations on classical computers have minor contributions to the overall time. 
\begin{remark}[Difference between the real space and the Fourier space readouts]
In principle, the readout in the real space is efficient for localized functions, that is, the number of grid points with nonzero values is sufficiently small compared to the grid number $N$. A best example is the linear combination of Dirac delta functions where the number of nonzero values is independent of $N$. 
On the other hand, the readout in the Fourier space addresses the dominant Fourier coefficients so that the reconstruction result yields the ``shape" of the function provided that it is sufficiently smooth. The best examples are smooth periodic functions for which the parameter $s$ is nearly zero. 
\end{remark}
\begin{remark}[Quantum advantages in CAE simulations]
\label{sec3:rem3}
In the CAE simulation, the most expensive step is the solution of a PDE. Previous works on the quantum algorithms for solving the PDEs show that the solutions encoded in the quantum states can be prepared exponentially/polynomially faster than the conventional classical algorithms (e.g., sparse conjugate gradient method \cite{Shewchuk1994}), that is, $\mathcal{C}_\psi=O(\mathrm{polylog} N)$ or $O(\mathrm{poly}\, N^{1/d})$. This implies that quantum algorithms help us to extend the tractable problem size. 
On the other hand, for the CAE problems in which the solutions themselves (instead of some physical quantities) are required, we have to include the readout of the solutions in the total cost. Table \ref{sec3:tab1} shows that the end-to-end quantum speedup in $N$ still survives for a fixed error bound. As for the high precision simulations, there is a hidden relation between the grid number and the error bound, e.g., $N=O(1/\varepsilon^d)$. Although the total computational cost of the quantum algorithms has limited acceleration compared to the classical algorithms in the worst scenario, we find that the computational time of the quantum algorithms, which is proportional to the maximal circuit depth, is much smaller if quantum devices can be used in parallel. 
\end{remark}

\subsection{Numerical examples}
\label{subsec:3-2}

We provide numerical examples to check the performance of the above methods. Provided that the grid number $N$ is sufficiently large, we demonstrate the decrease of the errors when the number of shots/queries to the oracle increases. 
The results in \cite[Table 1]{Patterson.2025} indicate that the QAE-based method in the real space is expensive when $J=N$. Thus, we omit the plot of the RSQAE method. The numerical results are obtained by the quantum simulator Qiskit \cite{Qiskit2023} without considering the quantum noises. Besides, we employ the RQAE algorithm in \cite{Manzano.2023} with the amplification policy $q=2$ and the confidence level $\gamma=0.05$ for the FSQAE method.   

\noindent \underline{Example 1}\ A 2D linear combination of Gaussian functions with $N_1=N_2=2^9=512$:
$$
f_1(x,y) = \mathrm{exp}\left(-25\left((x-0.65)^2+(y-0.65)^2\right)\right) + \mathrm{exp}\left(-16\left((x-0.35)^2+(y-0.35)^2\right)\right),
$$
for $(x,y)\in [0,1]^2$. 
\begin{figure}
\centering
\resizebox{14cm}{!}{
\includegraphics[keepaspectratio]{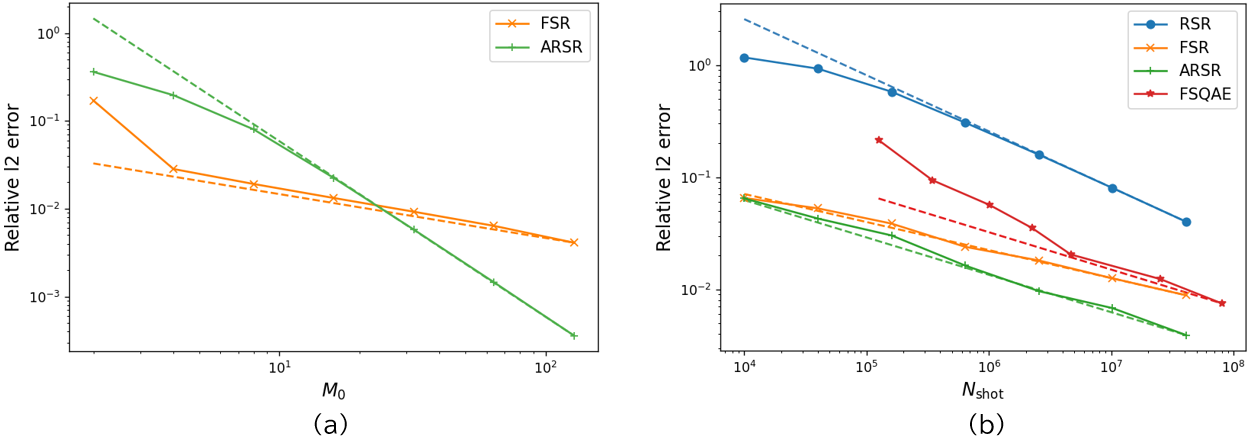}
}
\caption{Error plots for different methods in the example of a 2D linear combination of Gaussian functions. (a) Approximation errors (i.e., $N_{\text{shot}}\to \infty$) regarding $M_0\in \{2^1,2^2,\ldots,2^7\}$. The dashed lines denote the orders $O\left(1/M_0^{t}\right)$ with $t=1/2,2$ for the FSR and the ARSR methods, respectively. (b) Total errors regarding the number of shots $N_{\text{shot}}\in 10000\times \{4^0,4^1,\ldots,4^6\}$. For the FSQAE method, the $x$-axis denotes the number of queries to the oracle, and the plot is done by taking $\epsilon_0\in \{0.05, 0.02, 0.01, 0.005, 0.0025, 0.001, 0.0005\}$ in the RQAE algorithm. The dashed lines denote the orders $O\left(1/N_{\text{shot}}^{t}\right)$ with $t=1/2,1/4,1/3,1/3$ for the RSR, the FSR, the ARSR, and the FSQAE methods, respectively. }
\label{sec3:fig1}
\end{figure}
We demonstrate the error plots for different methods in Fig.~\ref{sec3:fig1}. The dashed lines are reference lines indicating the theoretical orders in Table \ref{sec3:tab1} with $s=2/(2\times 1-1)=2$ for the FSR and $s=2/(2\times 2-1)=2/3$ for the FSQAE with the even extension. 
Here, the FSR method denotes the modified FSR using the quantum circuits in Appendix \ref{sec:appA} without the extension operator. The approximation parameters $M_1=M_2=M_0$ are chosen to minimize the relative $l^2$ errors between the true solution and the approximate solutions for the ARSR and the FSQAE methods. 
Figure \ref{sec3:fig1}(b) shows that the numerical orders well fit the theoretical orders. Although the decay of the RSR method seems better than the other methods, its error also depends on the grid number $N$ (see e.g., Figs.~\ref{appC2:fig3},\ref{appC2:fig4}), which is the main drawback for the RSR method.  
In this example, the ARSR method seems to be the best one with the order $N_{\text{shot}}=O\left(1/\varepsilon^{3}\right)$. 
Besides, we focus on the approximation error plotted in Fig.~\ref{sec3:fig1}(a) and find that the order of the FSR method is worse than the ARSR method. Although the function $f_1$ itself is smooth, the values do not coincide on the left and right (or the upper and bottom) boundaries. 
The approximation error of the FSR method decreases slowly as the theoretical order $O(1/M_0^{1/2p})=O\left(1/M_0^{1/2}\right)$ in this worst scenario (a general non-periodic function). On the other hand, for a sufficiently smooth function, we can always apply the even extension so that the left and right (and the upper and bottom) boundary values coincide. This implies that the parameter $s$ for the FSR method is essentially smaller than $2/3$ as long as the function is continuous and sufficiently smooth (i.e., piecewise $W^{2,1}$). 

\begin{figure}
\centering
\resizebox{14cm}{!}{
\includegraphics[keepaspectratio]{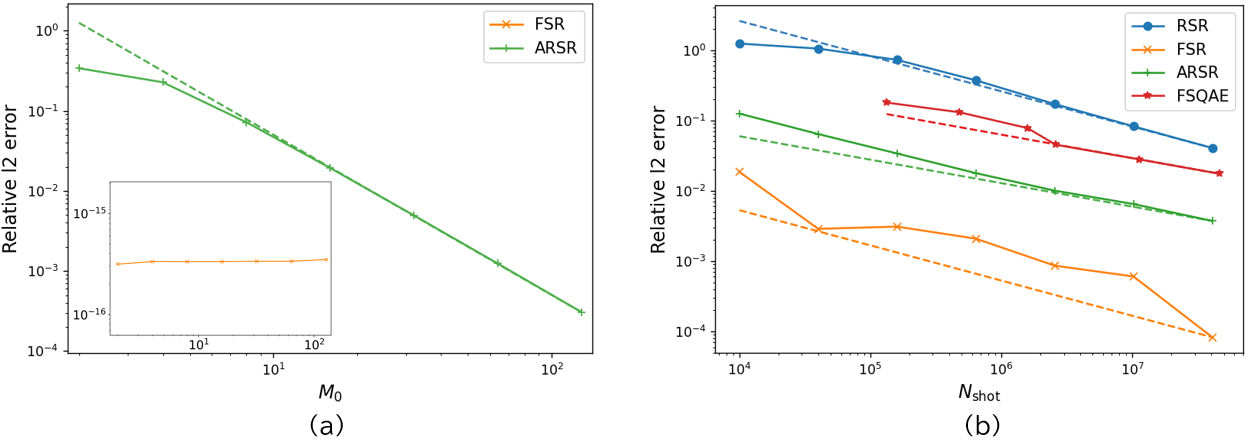}
}
\caption{Error plots for different methods in the example of a 2D trigonometric function. (a) Approximation errors (i.e., $N_{\text{shot}}\to \infty$) regarding $M_0\in \{2^1,2^2,\ldots,2^7\}$. The dashed line denotes the order $O\left(1/M_0^{t}\right)$ with $t=2$ for the ARSR method. (b) Total errors regarding number of shots $N_{\text{shot}}\in 10000\times \{4^0,4^1,\ldots,4^6\}$. For the FSQAE method, the $x$-axis denotes the number of queries to the oracle, and the plot is done by taking $\epsilon_0\in \{0.05, 0.02, 0.01, 0.005, 0.0025, 0.001\}$ in the RQAE algorithm. The dashed lines denote the orders $O\left(1/N_{\text{shot}}^{t}\right)$ with $t=1/2,1/2,1/3,1/3$ for the RSR, the FSR, the ARSR, and the FSQAE methods, respectively. }
\label{sec3:fig2}
\end{figure}

\noindent \underline{Example 2}\ A shifted 2D sine function with $N_1=N_2=2^9=512$:
$$
f_2(x,y) = \sin(2\pi x)\sin(2\pi y) + 1, \quad (x,y)\in [0,1]^2. 
$$
Again, we demonstrate the error plots for different methods in Fig.~\ref{sec3:fig2} with the reference lines indicating the theoretical orders in Table \ref{sec3:tab1} with $s=2/(2\times \infty-1)=0$ for the FSR and $s=2/(2\times 2-1)=2/3$ for the FSQAE with the even extension. 
Figure \ref{sec3:fig2}(b) shows that the numerical orders fit the theoretical ones well. In this example, the FSR method is the best since the Fourier coefficients have an exponential decay regarding $M_0$, which implies that the integer $p$ in Sect.~\ref{subsec:3-1} is infinity. Thus, $s=0$ and $N_{\text{shot}}=O\left(1/\varepsilon^{2}\right)$ is achieved without any dependence on $N$. 
This shows that the FSR method is remarkable for smooth periodic functions. 
On the other hand, although the function is smooth, its even extension, whose derivative is not continuous, has limited regularity at the extended boundaries. This is why the parameter $s$ for the FSQAE cannot be strictly smaller than $2/3$. 
However, for such a smooth periodic function, we can avoid the function extension and combine the QAE algorithm with the idea of the modified FSR method to derive the real and imaginary parts of the Fourier coefficients. This yields the best asymptotic order $O(1/\varepsilon)$, see Appendix \ref{subsec:C2-3}. 

\subsection{Features of methods}
\label{subsec:3-3}

To sum up, we give remarks on the strong and weak points of the mentioned methods. 

The RSR method has good dependence on the error bound if the grid number $N$ is limited. However, this contradicts the fundamental quantum advantage regarding $N$. Thus, the RSR method is unsuitable for the readout of general functions, whereas it can still be useful for obtaining only the peak location of the localized functions.  

The ARSR method (or the RSR method with the RMS post-processing) eliminates the $N$-dependence in the RSR method at the cost of increasing the $\varepsilon$-dependence. 
A crucial problem is the practical criteria to determine the approximation parameters when limited a priori information of the function is known. Another problem is the efficient determination of the signs because the sampling-based method only gives the absolute value of the amplitude. Still, it is a good candidate when the function is known to be sufficiently smooth (e.g., $f\in C^2$) with a fixed sign.  

The FSR method reads out the Fourier coefficients instead of the values at the grid points. This also eliminates the $N$-dependence for general functions. One key advantage of the FSR method is the adaptive determination of the approximation parameters. This implies we can apply it to the real-valued functions without any a priori information. On the other hand, the efficiency highly relies on the decay rate of the Fourier coefficients, which depends on the boundary coincidence and the regularity of the function. Although it has a worse order than the ARSR method in the worst scenario, it performs better for larger dimensions and has distinguished performance for sufficiently smooth periodic functions. 

The FSQAE method combines the idea of the QAE and the readout in Fourier space. There is also no $N$-dependence in the quantum complexity, and it has the best asymptotic behavior when the parameter $s$, indicating the decay of the Fourier coefficients, is smaller than $\min\{d/2, 2/(3d-2)\}$. Similar to the ARSR method, a crucial problem is the determination of the approximation parameters. Although the large maximal depth of the quantum circuits and large pre-factor limit its practical applications on near-term quantum devices, it will be the best approach among the mentioned methods in the FTQC era for sufficiently smooth periodic functions when high precision is required. 

Despite their simplicity, the ARSR and the FSR methods are among the best candidates for general continuous real-valued functions on near-term/mid-term quantum devices. 
In Appendix \ref{sec:appF}, we provide rough estimations of the required numbers of shots for the RSR, the ARSR, and the FSR methods in practical situations when $s=2/3$, as well as the worst situations when the underlying functions are discontinuous. 

\section{Applications to CFD simulations}
\label{sec:4}

We first investigate the visualization of the quantum solutions in some simple but practical problems. In addition, we show the results of solving a (nonlinear) Burgers' equation without the linearization strategy (e.g., Carleman linearization \cite{Liu.2021, Krovi2023}). 

\subsection{Visualization of CFD solutions}
\label{subsec:4-1}

We apply the RSR, the FSR, and the ARSR methods to visualize some CFD solutions in two representative flows. In this section, we use the modified FSR method in Appendix \ref{sec:appA} as the FSR method.

\begin{figure}[htp]
\centering
\resizebox{14cm}{!}{
\includegraphics[keepaspectratio]{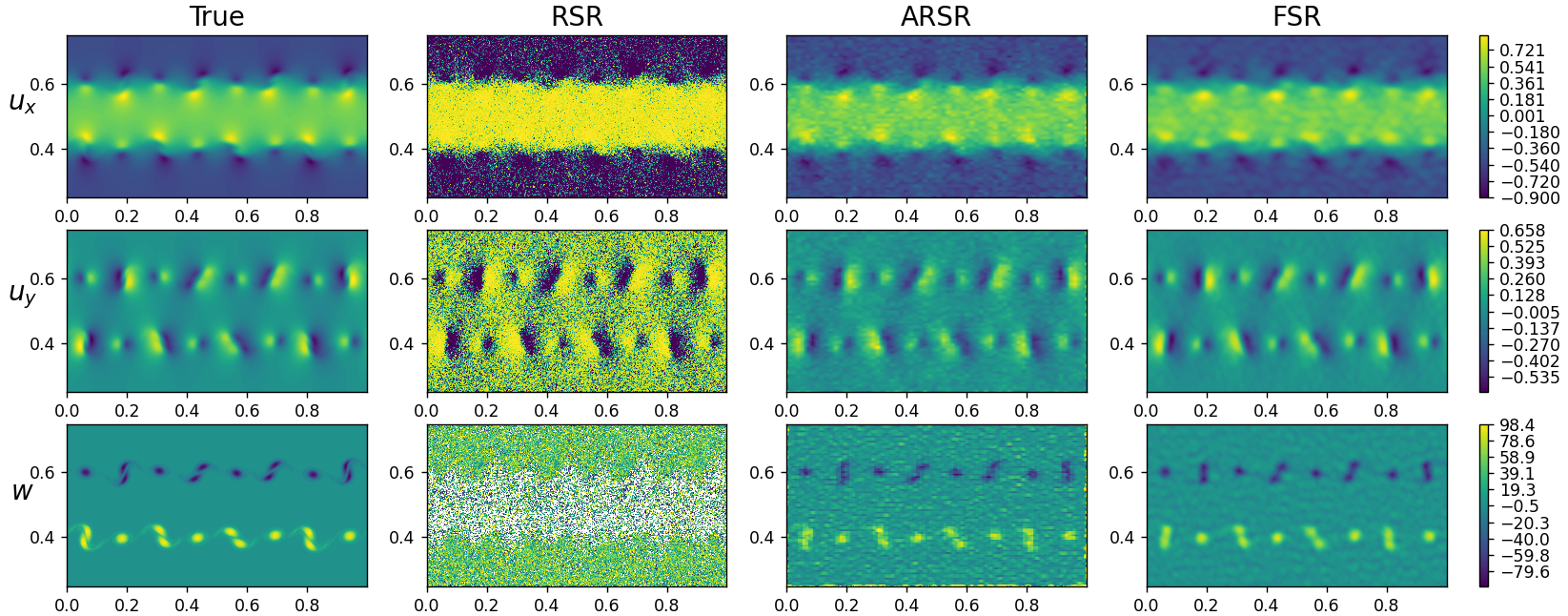}
}
\caption{Visualization of solution for a planar jet flow and comparison of different methods with $N_{\text{shot}}=1.6\times 10^5$. The 2D curls are calculated using the (reconstructed) velocity fields.} 
\label{sec4:fig1}
\end{figure}

\begin{figure}[htp]
\centering
\resizebox{14cm}{!}{
\includegraphics[keepaspectratio]{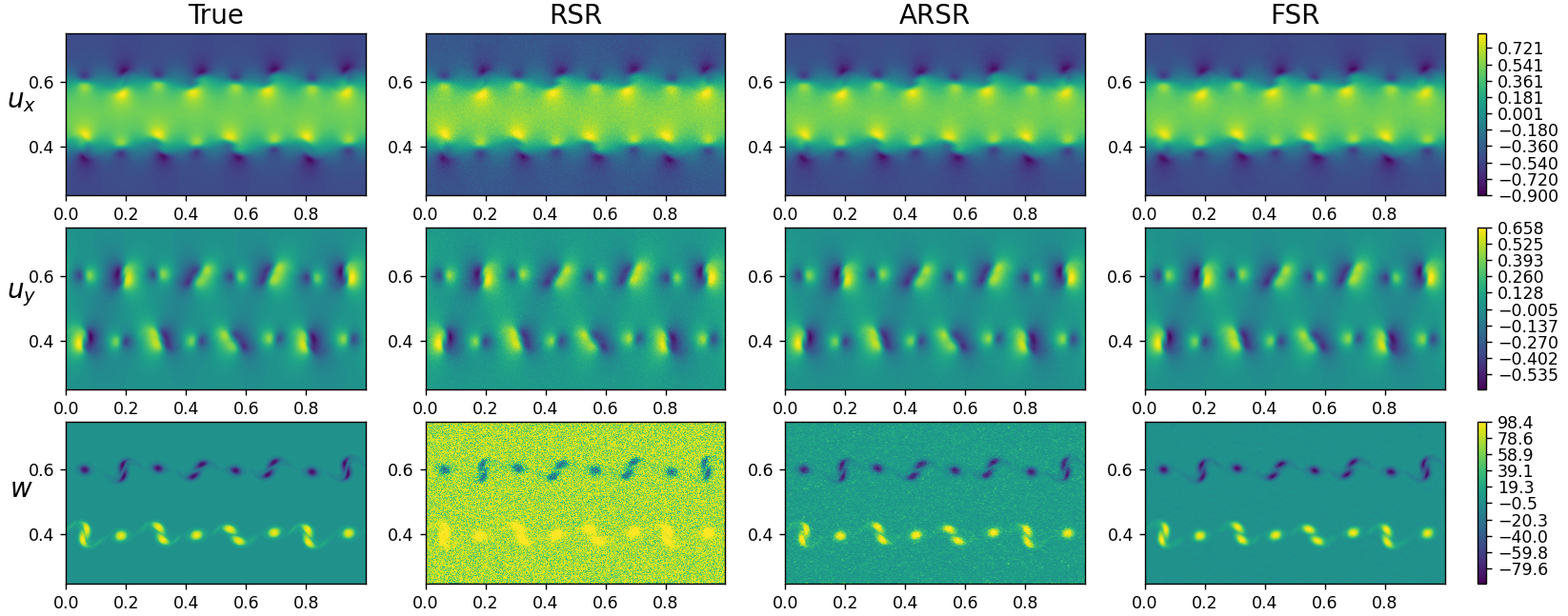}
}
\caption{Visualization of solution for a planar jet flow and comparison of different methods with $N_{\text{shot}}=4.096\times 10^7$. }
\label{sec4:fig2}
\end{figure}

\noindent \underline{2D temporally developing jet} The first example is a planar jet streaming along the $x$-direction with Reynolds number $Re=1000$, which was discussed in \cite{Gourianov.2022}. 
The authors used the tensor network approach to solve the incompressible Navier-Stokes equations. Here, we assume that the solution at a fixed time has already been prepared in a given quantum state and its $l^2$ norm is known (e.g., by the success probability of the quantum solver). 
We borrow the $2^{10}\times 2^{10}$ solution (i.e., velocity field $\mathbf{u}=(u_x, u_y)$) at $t=0.75$ from Code Ocean \cite{Gourianov.2021code} as the ``true" solution and use the Qiskit function $\mathbf{set\_statevector()}$ to precisely prepare the oracle $U_\psi$. 
The readouts of the velocity field by different methods are illustrated in Fig.~\ref{sec4:fig1} with relatively small $N_{\text{shot}}=1.6\times 10^5$, and in Fig.~\ref{sec4:fig2} with relatively large $N_{\text{shot}}=4.096\times 10^7$. 
Moreover, the 2D curls, e.g., $w:= \partial_x u_y - \partial_y u_x$, are calculated by the nine-point stencil finite difference approximation in the third row. For the RSR and the ARSR methods, we assume additional knowledge of the minimums of the solutions, and that the shifted functions can be prepared by using the oracle $U_\psi$. Besides, we determine the approximation parameters for the ARSR method by the convergence criteria ($m_1=m_2$ starts from $4$) in Remark \ref{sec2:rem1}. 

\begin{figure}[htp]
\centering
\resizebox{13cm}{!}{
\includegraphics[keepaspectratio]{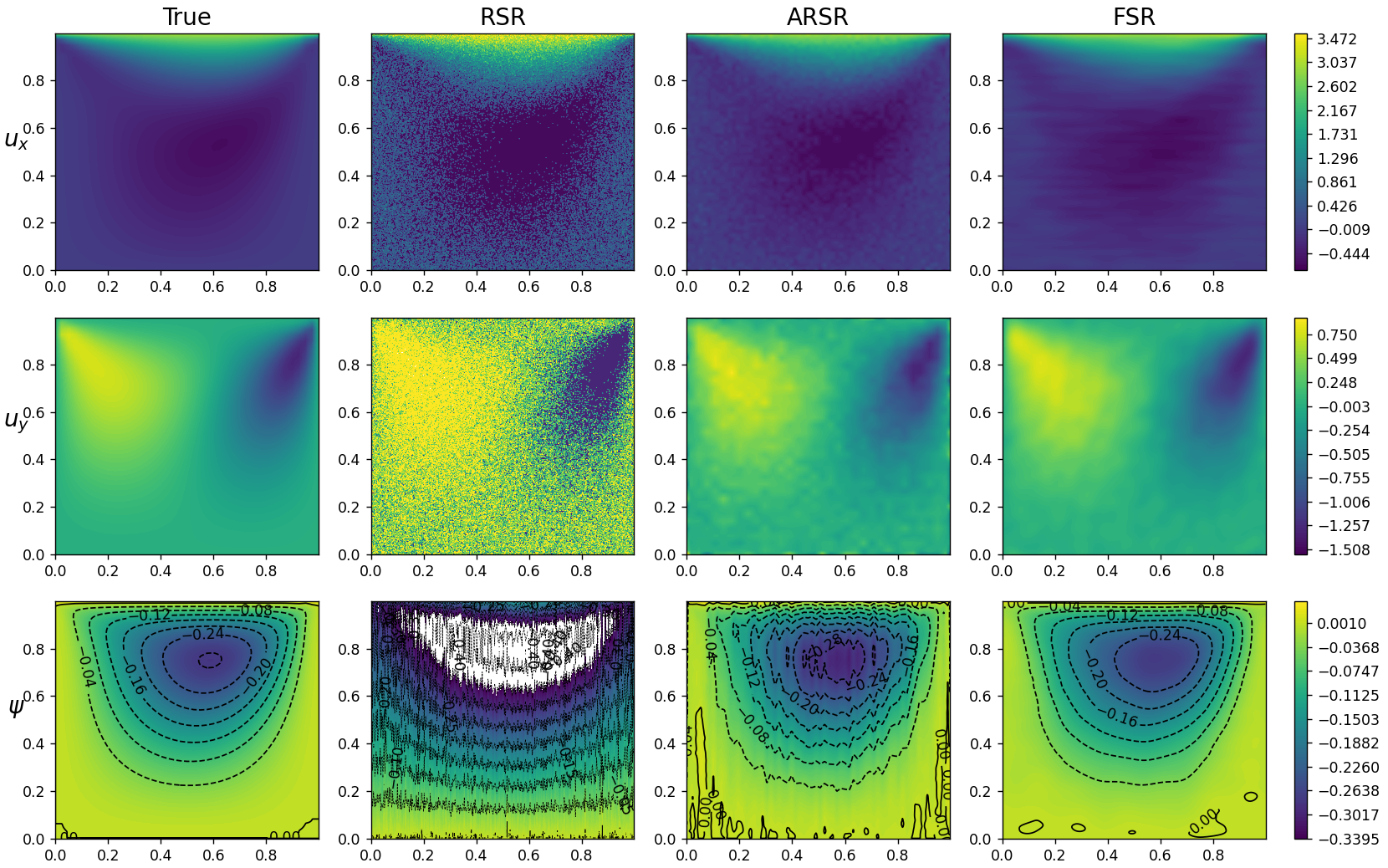}
}
\caption{Visualization of solution for a lid-driven cavity flow and comparison of different methods with $N_{\text{shot}}=1.6\times 10^5$. The plots of the stream function $\psi$ are obtained by numerical integration of the (reconstructed) velocity fields. }
\label{sec4:fig3}
\end{figure}

\begin{figure}[htp]
\centering
\resizebox{13cm}{!}{
\includegraphics[keepaspectratio]{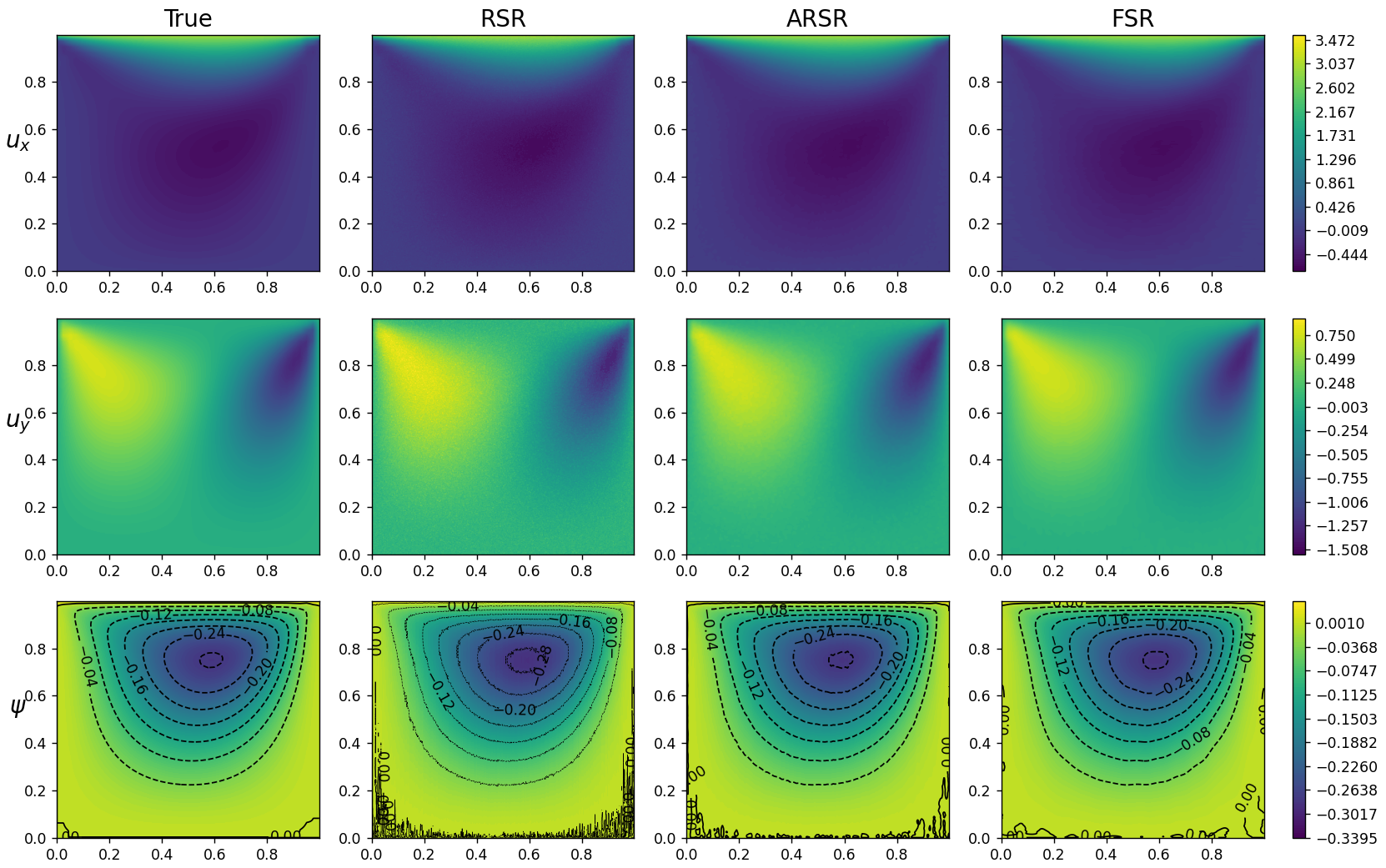}
}
\caption{Visualization of solution for a lid-driven cavity flow and comparison of different methods with $N_{\text{shot}}=4.096\times 10^7$. }
\label{sec4:fig4}
\end{figure}

\noindent \underline{2D lid-driven cavity flow} The second example is a benchmark problem for a cavity flow with non-zero speed in $x$-direction on the top boundary. 
We directly borrow a steady state velocity field $\mathbf{u}=(u_x,u_y)$ from the datasets \cite{Burkardt.2006data}. The original data are $41\times 41$ matrices, but we use the spline interpolation to generate the velocity field with a larger grid number $N=2^9\times 2^9$.  
The results by different methods are illustrated in Fig.~\ref{sec4:fig3} using relatively small $N_{\text{shot}}=1.6\times 10^5$, and in Fig.~\ref{sec4:fig4} using relatively large $N_{\text{shot}}=4.096\times 10^7$. 
In the third row, the stream functions, e.g., $\psi$ satisfying $u_x=\partial_y\psi$ and $u_y=-\partial_x\psi$, are calculated by the numerical integration of $u_x$ (possibly varying up to a constant). 
Moreover, for the FSR method, we extend the solution $u_x$ in the $y$-direction since it is non-periodic. This can be done by a similar extension operator in \cite{Huang.2025pre} with additional gates to modify the values at the top boundary.  

According to Figs.~\ref{sec4:fig1}--\ref{sec4:fig4}, the FSR method generates the shape of the solution quickly and greatly outperforms the RSR method when the number of shots is limited. On the other hand, if we have the a priori information of the minimal values of $u_x$ and $u_y$, then it is possible to use a similar circuit as Fig.~\ref{appB:fig1} in Appendix \ref{sec:appB} to shift the amplitudes so that they become non-negative. In such cases, the ARSR method yields similar results to the FSR method. 
In both examples, the functions are not periodic even though the left and right (or the top and bottom) boundaries have the same values. One can extend the functions beyond the boundary to make them periodic, but the regularities (i.e., smoothness) of the extended functions are limited. Therefore, the parameter $s$ for the FSR method is $2/3$ (see also Appendix \ref{subsec:C2-2} for the numerical confirmation), which results in a similar performance to the ARSR method. 

\subsection{Path to time stepwise readout strategy for solving evolution equations}
\label{subsec:4-2}

The efficient ($N$-independent) readout methods enable the time stepwise readout (TSR) strategy for solving evolution equations on quantum computers. 
We demonstrate the idea using the 2D Burgers' equation:
$$
\partial_t \mathbf{u}(\mathbf{x}, t) = \nu \nabla^2 \mathbf{u}(\mathbf{x}, t) - \mathbf{u}\cdot\nabla \mathbf{u}(\mathbf{x}, t), \quad \mathbf{x}=(x,y)\in [0,2\pi]^2, \ t>0,
$$
with the periodic boundary condition and the viscosity $\nu=0.05$. In addition, we take the initial conditions as follows:
$$
u_x^{(0)}(x,y) := C\sin(x+y),\quad u_y^{(0)}(x,y) := C\sin(x+y), \quad (x,y)\in [0,2\pi]^2, 
$$
where $C=1/(2\pi)$ is a constant such that the initial conditions are normalized: $\|\mathbf{u}^{(0)}\|=1$. 
 
To solve a nonlinear evolution equation, one explicit approach is based on the time marching method. In other words, we integrate the equation for a sufficiently small time interval $\Delta t$ in each time step and use the solution in the current time step to approximate the nonlinear term in the next time step. 
As for the quantum implementation, although the improved time-marching based quantum solvers were developed in \cite{Fang.2023}, the block encoding of the matrix and the QAA algorithm make it hard to realize on near-term/mid-term quantum devices. Therefore, we avoid the block encoding of the matrices and choose to use the efficient PITE algorithm proposed in \cite{Huang.2024pre} as the time marching method. For a nonlinear equation, the key difference from its linear counterpart is that the readout of the solution is indispensable in each time step, which was extremely expensive with the naive readout methods in real space (e.g., the RSR method). Using the $N$-independent readout method (e.g., the FSR method), we propose the PITE-TSR strategy to solve the nonlinear equations as follows: 
\begin{enumerate}[1)]
\item In the 1st time step, we prepare the initial quantum state corresponding to the initial conditions and apply the PITE operators for one time step, where the nonlinear term is implemented using the initial conditions $\mathbf{u}^{(0)}$. After that, we post-select the ancillary qubit for the PITE algorithm to be $\ket{0}$ and employ the FSR method to reconstruct the approximate solutions at $t=\Delta t$: $\mathbf{\tilde u}^{(1)}$. 

\item In the $k$-th time step for $k\ge 2$, we prepare the quantum state corresponding to the approximate solutions $\mathbf{\tilde u}^{(k-1)}$ and apply the PITE operators for one time step where the nonlinear term is implemented using the approximate solutions $\mathbf{\tilde u}^{(k-1)}$. Again, we post-select the ancillary qubit for the PITE algorithm to be $\ket{0}$ and employ the FSR method to reconstruct the approximate solutions at $t=k\Delta t$: $\mathbf{\tilde u}^{(k)}$.

\item We repeat the above step until $k=K$ for some integer $K$. Finally, we obtain the approximate solutions at $t=T:=K\Delta t$.  
\end{enumerate}
The circuit implementation and complexity of the above strategy are discussed in Appendix \ref{sec:appD}. To avoid confusion, we mention the difference of the proposed TSR strategy from the ones in \cite{Ye.2024, Song.2025, Zhuang.2025pre} where the variational algorithm VQE/VQLS or the HHL algorithm was applied in each time step. For large grid numbers, the variational algorithms meet the difficulty in optimization, while the HHL algorithms (or the QLSAs) meet the problem of large condition numbers. The approximate PITE algorithm \cite{Huang.2024pre} used in the proposed TSR strategy avoids both problems, and thus is adopted. 

\begin{figure}[tp]
\centering
\resizebox{14cm}{!}{
\includegraphics[keepaspectratio]{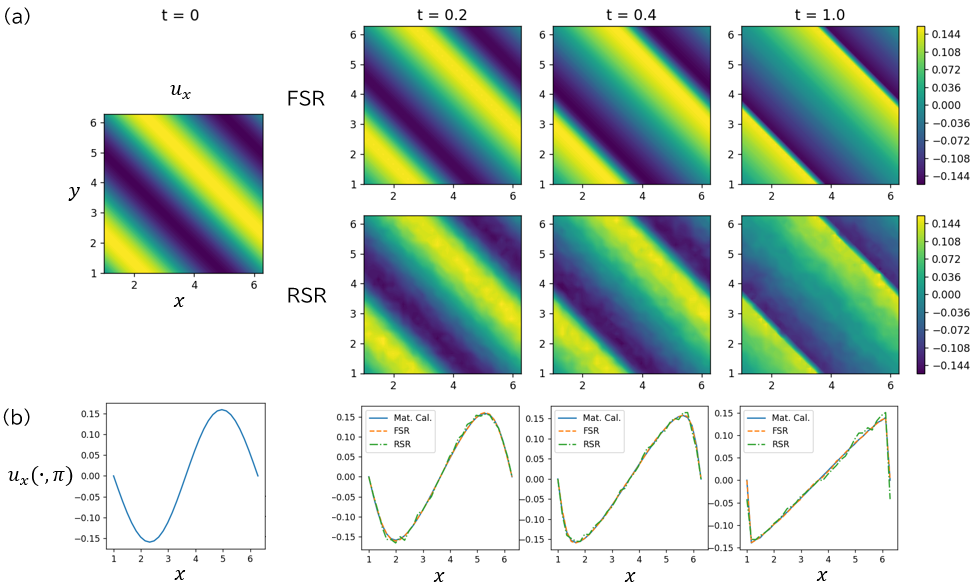}
}
\caption{Simulation results for the velocity in $x$-direction at different time $t=0.2, 0.4, 1.0$. (a) Quantum solutions using the FSR method and the RSR method. 
(b) Solutions that project to a line $y=\pi$. The blue lines denote the reference solutions using the matrix multiplications. }
\label{sec4:fig5}
\end{figure}

Here, we show the numerical solution $u_x$ to the above Burgers' equation in Fig.~\ref{sec4:fig5}, while we omit the similar result on $u_y$. 
We take $N=2^5\times 2^5=1024$ and $\Delta t=0.04$ so that the discretization error and the approximation error in the PITE algorithm are minor compared to the error introduced by the readout. The number of shots in each time step is about $1\times 10^6$, and the detailed setting of the simulation is provided in Appendix \ref{subsec:D-2}. 
The evolution from a sine wave to a sawtooth wave is demonstrated as expected. In fact, the relative $l^2$ error between the reference solution and the quantum solution using the (modified) FSR method is $0.019$, while the relative $l^2$ error is $0.128$ for the RSR method. 
In this example with the periodic boundary condition, the FSR method outperforms the RSR method even for such a small grid number $N=1024$. Besides, we omit the plots for the ARSR method because it is worse than the RSR method owing to the small grid number. 
Although the result by the RSR method in Fig.~\ref{sec4:fig5} seems to capture the rough ``shape" of the solution, its relative $l^2$ error has a linear scaling in $N$ and thus, the simulation would fail for a larger grid number (e.g., $N=2^7\times 2^7=16384$). In contrast, the FSR method would still provide a good solution for large grid numbers since its error is independent of $N$. 

In this paper, we do not numerically compare the proposed TSR strategy with the other existing quantum algorithms because the previous algorithms without the time stepwise readout, to our best knowledge, are hard to simulate at the gate level for this 2D nonlinear equation. For example, the well-known algorithm by the Carleman linearization \cite{Krovi2023} requires a large amount of qubits and large circuit depth for this problem, which is not practical for near-term/mid-term quantum devices. 
We end this section with a remark on the key strong point of the proposed TSR strategy. 
\begin{remark}[$K$-uniform success probability]
\label{sec4:rem4}
It was pointed out in \cite{Fang.2023} that the quantum algorithm for the evolution equations may suffer from a fast decay of success probability as the number of time step $K$ increases according to the implementation of the matrix in each time step (e.g., the effect of sub-normalization in the block encoding of matrices \cite{Camps.2024, Sunderhauf.2024}). This usually occurs for non-unitary operations and yields an exponentially small total success probability $O(c_0^K)$ for some constant $c_0<1$. 
While one can apply the expensive QAA algorithm to enlarge the success probability in each time step, the TSR strategy avoids such a crucial problem without employing the QAA algorithm. This greatly reduces the maximal circuit depth. 
Moreover, in each time step, the success probability is uniform regarding $k\le K$, and thus the quantum complexity has a linear scaling regarding $K$ instead of an exponential one. For example, the success probability in each time step has a lower bound $0.3$ with the TSR strategy, while the total success probability is smaller than $1\times 10^{-7}$ without the TSR strategy for the simulation of the above 2D Burgers' equation. 
\end{remark}


\section{Conclusion}
\label{sec:5}

We are devoted to general readout methods for real-valued functions without optimization in either real or Fourier space. In addition to the conventional real space readout (RSR) and the Fourier space readout (FSR) \cite{Huang.2025pre} methods, we proposed an efficient approximate real space readout (ARSR) method as well as the QAE-based methods: RSQAE and FSQAE. 

We compared the above methods and summarized their quantum and classical complexities. According to the theoretical upper bound and the numerical examples for some test functions, we conclude that 
\begin{itemize}
\item The RSR method is efficient for small grid number $N$, while it has extremely poor performance for large $N$ and relatively small number of shots $N_{\text{shot}}$. 

\item For sufficiently large $N$, the ARSR method relieves the effect of the sampling error in the RSR method and has good performance for sufficiently smooth positive/negative functions. In particular, it can outperform the readout methods in the Fourier space if the function is non-periodic and the spatial dimension $d$ is smaller than $2$. 

\item The RSQAE method is good if function values at only limited target points are required. For example, the number of target grid points $J$ satisfies $J=O(\log N)$. 

\item The FSR method works regardless of the grid number and is efficient for large grid numbers. Its efficiency highly depends on the regularity of the function and the boundary coincidence. In particular, it has a remarkable performance if the function is smooth and periodic. 

\item The FSQAE method combines the QAE algorithm with the readout in Fourier space. Although it is not efficient for a relatively large error bound, it has the best asymptotic behavior in the error bound provided that the function is sufficiently good (see the second last paragraph in Sect. \ref{sec:3}).

\end{itemize}
The ARSR, the FSR, and the FSQAE methods have $O(\mathrm{polylog}N)$ dependence in their quantum complexities, which indicates that the readout of functions in the CAE problems can be efficient regarding the grid number $N$. In spite of their simplicity, the ARSR and the FSR methods appear to be the currently best readout methods for continuous real-valued functions in practical problems. 

Moreover, we applied the readout methods to visualize the solutions in the CFD simulations of a planar jet flow and a lid-driven cavity flow. Compared to the RSR method, the FSR and the ARSR methods exhibit better results as the functions can be approximately reconstructed under a relatively small number of shots (see Figs.~\ref{sec4:fig1},\ref{sec4:fig3}). The numerical performance indicates that the regularity parameter in the FSR method is $s=2/3$ for the benchmark solutions in the CFD simulations, which has a slightly better performance than the ARSR method. As for high-precision simulations of complicated flows, it seems that the currently best methods are still costly.

Furthermore, we validated the time stepwise readout (TSR) strategy for solving the time evolution equations. More precisely, we combined the readout methods with the approximate PITE algorithm in \cite{Huang.2024pre} to solve a (nonlinear) 2D Burgers' equation. With the help of efficient readout methods whose quantum complexity is independent of the grid number, the proposed TSR strategy gives a practical and efficient algorithm to solve nonlinear evolution equations. 
As discussed in Remark \ref{sec4:rem4}, the TSR strategy avoids the fast decay of the success probability regarding the number of time steps, which leads to a smaller total quantum complexity.  

Finally, we mention several future topics. Although we have extensively compared the readout methods in the real-grid and the Fourier bases, we do not touch the readout methods in other orthonormal bases, including the Chebyshev basis. Despite the promising performance of the Chebyshev basis, one has to discuss the efficient circuit implementation and carefully investigate the success probability since the Chebyshev transform is non-unitary. Some preliminary discussion has been done in \cite{Williams.2023pre}. 
Another important topic is the rigorous comparison between the proposed TSR strategy and the existing quantum algorithms for solving nonlinear evolution equations. In addition to the theoretical upper bound of the asymptotic complexity, the practical resource estimations for well-known CAE problems are also indispensable. 

\section*{Acknowledgments}

This work was partially supported by the Center of Innovations for Sustainable Quantum AI (JST Grant number JPMJPF2221).

\appendix

\section{A modified FSR method}
\label{sec:appA}

In this section, we introduce a modified quantum circuit for the FSR method where we avoid the extension operator and directly estimate the complex Fourier coefficients. To distinguish from the original FSR method \cite{Huang.2025pre}, we denote the modified one by the FSR2 method here.

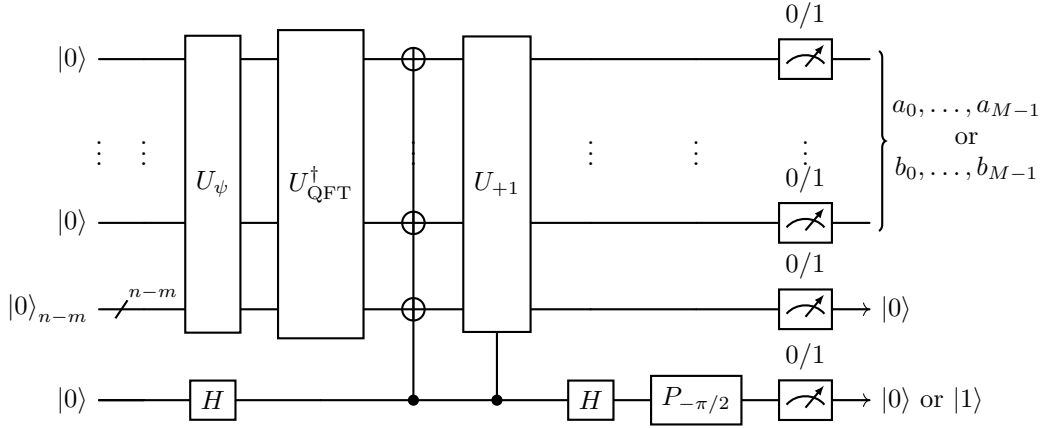
\begin{figure}
\centering
\begin{quantikz}
\lstick{\ket{0}} &  & \gate[4]{U_{\psi}} & \gate[4]{U_{\text{QFT}}^\dag} & \targ{} & \gate[4]{U_{+1}} &  &  & \meter{0/1} & \rstick[3]{$a_{0},\ldots,a_{M-1}$\\or \\$b_{0},\ldots,b_{M-1}$} \\[-0cm]
\setwiretype{n}\vdots & \vdots &  & \vdots & \vdots & \vdots &\vdots & \vdots & \vdots & \\
\lstick{\ket{0}} &  &  &  & \targ{} &  &  &  & \meter{0/1} &  \\[-0cm]
\lstick{$\ket{0}_{n-m}$} & \qwbundle{n-m} &  &  & \targ{} &  &  &  & \meter{0/1}\arrow[r] &
\rstick{\ket{0}} \\
\lstick{\ket{0}} & \qw & \gate{H} &  & \ctrl{-4} & \ctrl{-4} & \gate{H} & \gate{P_{-\pi/2}} & \meter{0/1}\arrow[r] &
\rstick{\ket{0} or \ket{1}} 
\end{quantikz}
\caption{Quantum circuit for determining the absolute values of the Fourier coefficients in the FSR2 method. By post-selecting the last qubit as $\ket{0}$, we repeat the measurements and obtain the absolute values of the real parts of the Fourier coefficients $a_j \approx |\mathrm{Re}(c_j)|$ for $j=0,\ldots,M-1$. We also obtain the absolute values of the imaginary parts of the Fourier coefficients $b_j \approx |\mathrm{Im}(c_j)|$ for $j=0,\ldots,M-1$ when the last qubit is $\ket{1}$.}
\label{appA:fig1}
\end{figure}

\begin{figure}
\centering
\begin{quantikz}
\lstick{$\ket{0}_{m}$} & \qwbundle{m} &  & \gate[2]{U_{\psi}} & \gate[2]{U_{\text{QFT}}^\dag} & \targ{} & \gate[2]{U_{+1}} &  &  & \gate{H^{\otimes m}} &  & \meter{0/1} & \\[-0cm]
\lstick{$\ket{0}_{n-m}$} & \qwbundle{n-m} &  &  &  & \targ{} &  &  &  &  &  & \meter{0/1} & \\
\lstick{\ket{0}} & \qw & \gate{H} &  &  & \ctrl{-2} & \ctrl{-2} & \gate{H} & \gate{P_{-\pi/2}} & \gate{H} &  & \meter{0/1}\arrow[r] &
\rstick{\ket{q}} \\
\lstick{\ket{0}} & \qw & \gate{H} & \ctrl{-2} & \ctrl{-2} & \ctrl{-3} & \ctrl{-2} &  & \ctrl{-1} & \octrl{-3} & \gate{H} & \meter{0/1}\arrow[r] & \rstick{\ket{0}}
\end{quantikz}
\caption{Quantum circuit for determining the signs of the Fourier coefficients in the FSR2 method. By post-selecting the last qubit as $\ket{0}$ and the second last qubit as $\ket{q}$, we obtain $e_j^{\text{re}} \approx \frac12 |\mathrm{Re}(c_j)+1/\sqrt{2M}|$ while $q=0$ and $e_j^{\text{im}} \approx \frac12 |\mathrm{Im}(c_j)+1/\sqrt{2M}|$ while $q=1$ for $j=0,\ldots,M-1$.}
\label{appA:fig2}
\end{figure}
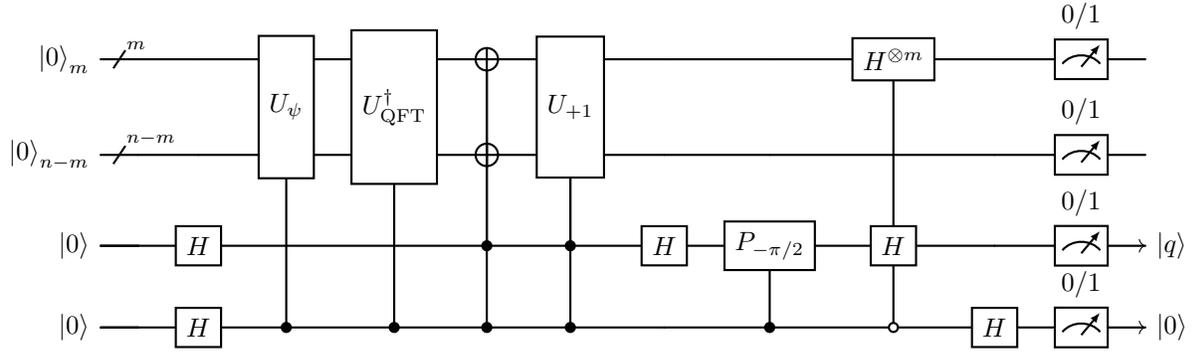

For the 1D case, the quantum circuit is given in Fig.~\ref{appA:fig1}. 
We can introduce the auxiliary quantum circuit for determining the signs of the real/imaginary parts of the Fourier coefficients as follows, see Fig.~\ref{appA:fig2}. By post-selecting the last qubit as $\ket{0}$ and the second last qubit as $\ket{0}$ or $\ket{1}$, we obtain $e_j^{\text{re}} \approx \frac12 |\mathrm{Re}(c_j)+1/\sqrt{2M}|$ or $e_j^{\text{im}} \approx \frac12 |\mathrm{Im}(c_j)+1/\sqrt{2M}|$ for $j=0,\ldots,M-1$. Together with the approximate absolute values of the real/imaginary parts of the Fourier coefficients that are obtained by the circuit in Fig.~\ref{appA:fig1}, we can determine the signs using the criterion in \cite[Section 2.3]{Huang.2025pre}. In this paper, we apply a more efficient criterion similar to that in \cite[Eq.~(21)]{Dewitte.2025}. That is, we use the results when the last qubit is observed as $\ket{1}$ to obtain $\tilde e_j^{\text{re}} \approx \frac12 |\mathrm{Re}(c_j)-1/\sqrt{2M}|$ or $\tilde e_j^{\text{im}} \approx \frac12 |\mathrm{Im}(c_j)-1/\sqrt{2M}|$. Then, we have $\mathrm{Re}(c_j)/\sqrt{2M}\approx (e^{\mathrm{re}}_j)^2-(\tilde e^{\mathrm{re}}_j)^2$, and we can determine the signs by $\mathrm{sgn}(\mathrm{Re}(c_j)) \approx \mathrm{sgn}\left((e^{\mathrm{re}}_j)^2-(\tilde e^{\mathrm{re}}_j)^2\right)$, $j=0,\ldots,M-1$. The signs for the imaginary parts are similarly determined. 

Next, we consider the quantum circuits for the multi-dimensional FSR2 method. Here, we explain the idea in terms of a two-dimensional case. 
We note that the original function $f$ does not have a symmetric structure for each dimension. Then, by the definition of the (quantum) Fourier coefficients, we have $c_{k_1, k_2}=\overline{c_{N_1-k_1,N_2-k_2}}$ for $k_1=0,\ldots,N_1-1, k_2=0,\ldots,N_2-1$, but no clear relation between $c_{k_1, k_2}$ and $c_{k_1, N_2-k_2}$. This indicates that we need the Fourier coefficients $c_{k_1, k_2}$ for both $k_1 = 0,\ldots,M_1-1, k_2 = 0,\ldots, M_2-1$ and $k_1 = N_1-M_1,\ldots,N_1-1, k_2 = 0,\ldots,M_2-1$ to approximately reconstruct the function $f$ provided that $M_1,M_2$ are sufficiently large. To do this, it is indispensable to determine the signs of $4M_1M_2$ coefficients, that is, $\mathrm{Re}(c_{k_1,k_2}), \mathrm{Im}(c_{k_1,k_2})$ for $k_1 = 0,\ldots,M_1-1,N_1-M_1,\ldots,N_1-1, k_2 = 0,\ldots,M_2-1$. 
We demonstrate the quantum circuits in Figs.~\ref{appA:fig3},\ref{appA:fig4}. 
\begin{figure}
\centering
\begin{quantikz}
\lstick{$\ket{0}_{m_1}$} & \qwbundle{m_1} &  & \gate[4]{U_{\psi}} & \gate[2]{U_{\text{QFT}}^\dag} & \targ{} & \gate[2]{U_{+1}} &  &  &  & \meter{0/1} & \\[-0cm]
\lstick{$\ket{0}_{n_1-m_1}$} & \qwbundle{n_1-m_1} &  &  &  & \targ{} &  &  &  &  & \meter{0/1} & \\
\lstick{$\ket{0}_{m_2}$} & \qwbundle{m_2} &  &  & \gate[2]{U_{\text{QFT}}^\dag} & \targ{} &  & \gate[2]{U_{+1}} &  &  & \meter{0/1} & \\[-0cm]
\lstick{$\ket{0}_{n_2-m_2}$} & \qwbundle{n_2-m_2} &  &  &  & \targ{} &  &  &  &  & \meter{0/1} & \\
\lstick{\ket{0}} & \qw &  & \gate{H} &  & \ctrl{-4} & \ctrl{-4} & \ctrl{-1} & \gate{H} & \gate{P_{-\pi/2}} & \meter{0/1}\arrow[r] &
\rstick{\ket{q}} 
\end{quantikz}
\caption{Quantum circuit for determining the absolute values of the Fourier coefficients in the FSR2 method for a 2D case. By post-selecting the last qubit as $\ket{q}$, we obtain the absolute values of the real parts if $q=0$ and the absolute values of the imaginary parts if $q=1$.}
\label{appA:fig3}
\end{figure}
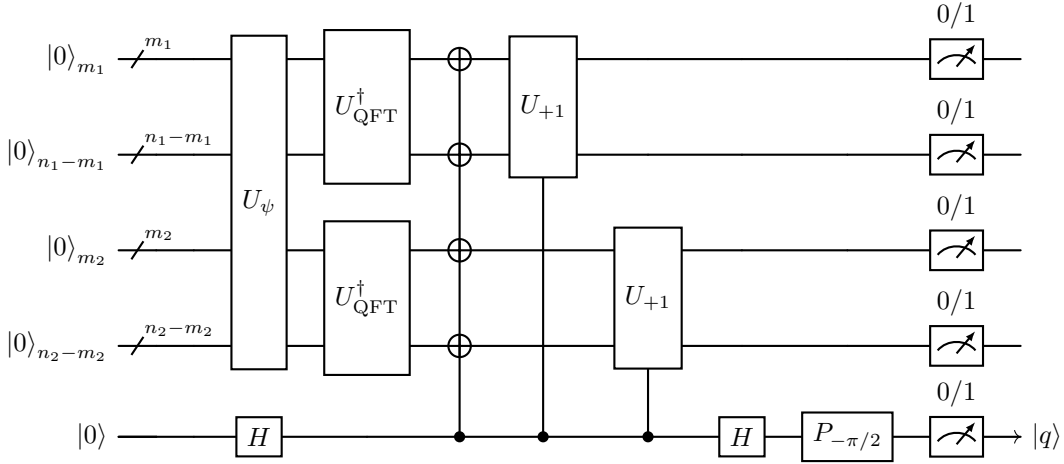
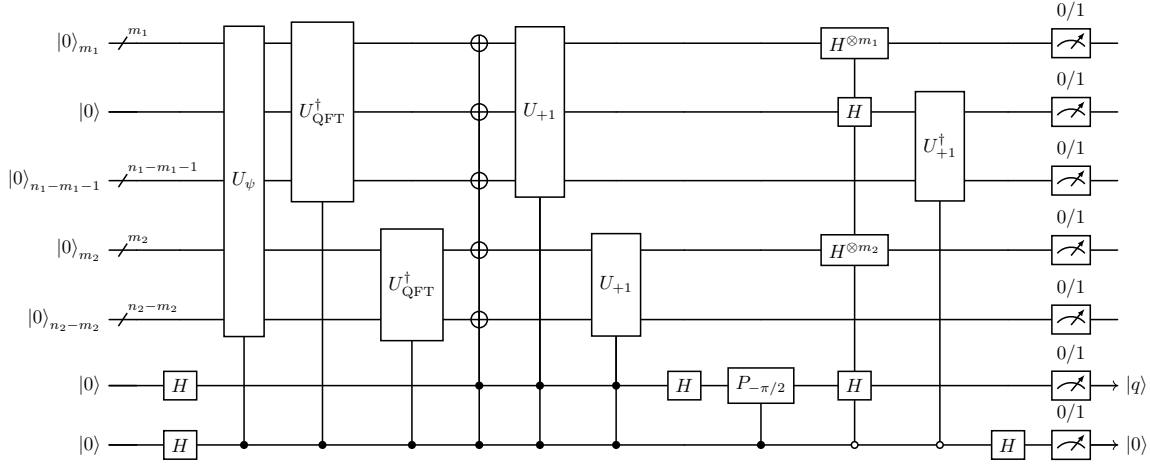
\begin{figure}
\centering
\resizebox{15.5cm}{!}{
\begin{quantikz}
\lstick{$\ket{0}_{m_1}$} & \qwbundle{m_1} &  & \gate[5]{U_{\psi}} & \gate[3]{U_{\text{QFT}}^\dag} &  & \targ{} & \gate[3]{U_{+1}} &  &  &  & \gate{H^{\otimes m_1}} &  &  & \meter{0/1} & \\[-0cm]
\lstick{\ket{0}} & \qw &  &  &  &  & \targ{} &  &  &  &  & \gate{H} & \gate[2]{U_{+1}^\dag} &  & \meter{0/1} & \\[-0cm]
\lstick{$\ket{0}_{n_1-m_1-1}$} & \qwbundle{n_1-m_1-1} &  &  &  &  & \targ{} &  &  &  &  &  &  &  & \meter{0/1} & \\
\lstick{$\ket{0}_{m_2}$} & \qwbundle{m_2} &  &  &  & \gate[2]{U_{\text{QFT}}^\dag} & \targ{} &  & \gate[2]{U_{+1}} &  &  & \gate{H^{\otimes m_2}} &  &  & \meter{0/1} & \\[-0cm]
\lstick{$\ket{0}_{n_2-m_2}$} & \qwbundle{n_2-m_2} &  &  &  &  & \targ{} &  &  &  &  &  &  &  & \meter{0/1} & \\
\lstick{\ket{0}} & \qw & \gate{H} &  &  &  & \ctrl{-4} & \ctrl{-4} & \ctrl{-1} & \gate{H} & \gate{P_{-\pi/2}} & \gate{H} &  &  & \meter{0/1}\arrow[r] & \rstick{\ket{q}} \\
\lstick{\ket{0}} & \qw & \gate{H} & \ctrl{-2} & \ctrl{-5} & \ctrl{-2} & \ctrl{-6} & \ctrl{-5} & \ctrl{-2} &  & \ctrl{-1} & \octrl{-6} & \octrl{-4} & \gate{H} & \meter{0/1}\arrow[r] & \rstick{\ket{0}} 
\end{quantikz}
}
\caption{Quantum circuit for determining the signs of the Fourier coefficients in the FSR2 method for a 2D case. By post-selecting the last qubit as $\ket{0}$ and the second last qubit as $\ket{q}$, we obtain $e_{j_1,j_2}^{\text{re}} \approx \frac12 |\mathrm{Re}(c_{j_1,j_2})+1/\sqrt{4M_1M_2}|$ while $q=0$ and $e_{j_1,j_2}^{\text{im}} \approx \frac12 |\mathrm{Im}(c_{j_1,j_2})+1/\sqrt{4M_1M_2}|$ while $q=1$ for $j_1=0,\ldots,M_1-1,N_1-M_1,\ldots,N_1-1$, $j_2=0,\ldots,M_2-1$.}
\label{appA:fig4}
\end{figure}
In Fig.~\ref{appA:fig4}, we introduce an inverse incrementer gate to shift the basis so that the constant $1/\sqrt{4M_1M_2}$ is added to the Fourier coefficients with indices $j_1=0,1,\ldots,M_1-1,N_1-M_1,\ldots,N_1-1$ and $j_2=0,\ldots,M_2-1$. Together with the result of the quantum circuit in Fig.~\ref{appA:fig3}, this helps us to determine the signs of the real/imaginary parts of the dominant Fourier coefficients. 
For general spatial dimension $d=1,2,\ldots$, one needs to determine the $2^{d-1}\prod_{\ell=1}^d M_\ell$ Fourier coefficients (with $2\times 2^{d-1}\prod_{\ell=1}^d M_\ell$ signs). Then, the controlled Hadamard gate and inverse incrementer gate should be applied to the quantum registers corresponding to the dimension indices $\ell=1,2,\ldots,d-1$, so that the signs of all the necessary Fourier coefficients can be estimated. 

As long as the approximate Fourier coefficients are obtained, we can approximately reconstruct the function by 
\begin{equation}
\label{appA:eq0}
f(\mathbf{x}) \approx C_N\sum_{k_1=-(M_1-1)}^{M_1-1}\cdots \sum_{k_d=-(M_d-1)}^{M_d-1} c_{\mathbf{k}} \prod_{\ell=1}^d \mathrm{exp}\left(\mathrm{i}\frac{2\pi}{L_\ell} k_\ell x_\ell\right), \quad \mathbf{x}\in [0,L_1] \times \cdots \times [0,L_d].
\end{equation}
Here, the constant $C_N=A_N/\sqrt{N}$ with $A_N := \left(\sum_{j=0}^{N-1}|f(\mathbf{x}_j)|^2\right)^{1/2}$, $\mathbf{x}_j = (j_1 L_1/N_1,\ldots,j_d L_d/N_d)$, and $N=\prod_{\ell=1}^d N_\ell$. 
Note that the negative indices of the Fourier coefficients are defined as follows:
\begin{equation}
\label{appA:eq1}
c_{\mathbf{k}} = c_{k_1,\ldots,k_d} := 
\left\{
\begin{aligned}
& c_{\tilde{k}_1,\ldots,\tilde{k}_{d-1}, k_d}, &&\quad k_d\ge 0, \\
& \overline{c_{N_1-\tilde{k}_1,\ldots,N_{d-1}-\tilde{k}_{d-1}, -k_d}}, &&\quad k_d<0,
\end{aligned}
\right.
\end{equation}
where $\tilde k_\ell$ is defined by
$$
\tilde{k}_\ell = 
\left\{
\begin{aligned}
& k_\ell, &&\quad k_\ell=0,\ldots,N_\ell-1, \\
& N_\ell + k_\ell, &&\quad k_\ell=-N_\ell,\ldots,-1.
\end{aligned}
\right.
$$
In the second line of Eq.~\eqref{appA:eq1}, the subindex $N_\ell-\tilde{k}_\ell$ becomes $N_\ell$ if $\tilde{k}_\ell=0$, which is out of range. However, owing to the definition of the Fourier coefficients, the subindex $N_\ell$ can be substituted by $0$. Therefore, we can reconstruct the function using the derived Fourier coefficients $c_{k_1,\ldots,k_d}$, $k_\ell = 0,\ldots,M_\ell-1, N_\ell-M_\ell,\ldots,N_\ell-1$, $\ell=1,\ldots,d-1$, $k_d=0,\ldots,M_d-1$. 
\begin{remark}
In the above discussion, we have applied a technique (see Eq.~\eqref{appA:eq1}) to halve the necessary Fourier coefficients according to the symmetric property of the Fourier coefficients. 
However, this introduces an additional approximation error if $c_{k_1,\ldots,k_{d-1},N_d/2}$ is not sufficiently small. 
In such a case, we can choose to estimate $2^d \prod_{\ell=1}^d M_\ell$ Fourier coefficients by modifying the quantum circuit in Fig.~\ref{appA:fig4} (i.e., adding the controlled Hadamard gate and inverse incrementer gate to the quantum registers corresponding to all the spatial dimensions). Then, the negative Fourier coefficients can be simply defined by
$$
c_{\mathbf{k}} = c_{k_1,\ldots,k_d} := c_{\tilde{k}_1,\ldots,\tilde{k}_d},
$$
instead of Eq.~\eqref{appA:eq1}, and we reconstruct the function by Eq.~\eqref{appA:eq0}.  
\end{remark}

\section{Quantum circuits for the QAE-based methods}
\label{sec:appB}

\begin{figure}
\centering
\begin{quantikz}
& \qwbundle{n} & \gate[2]{\text{$\mathcal{A}_j[b]$}} &  \\[-0cm]
& \qw &  & 
\end{quantikz}
= \quad
\begin{quantikz}
 &  & \gate[2]{U_\psi} & \gate[2]{U_{\text{MADD}}^\dag[j]} & \gate{R_y(\theta_b)} &  & \\[-0cm]
 & \qwbundle{n-1} &  &  &  &  & \\[-0cm]
 & \gate{H} & \ctrl{-1} & \ctrl{-1} & \octrl{-2} & \gate{H} &
\end{quantikz}
\caption{Quantum circuit for the oracle $\mathcal{A}_j[b]$. Here, $j\in \{0,\ldots,N-1\}$ and $b\in [-1,1]$ are given parameters, while $\theta_b:=2\arccos(b)$ is an angle calculated from $b$. Besides, $U_{\text{MADD}}^\dag[j]$ denotes the inverse modular adder, which is a modular subtractor.}
\label{appB:fig1}
\end{figure}
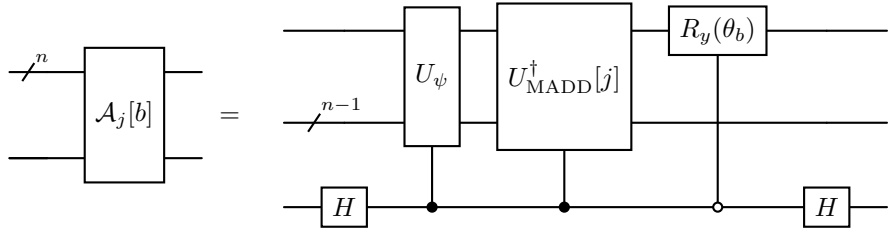

\begin{figure}
\centering
\begin{quantikz}
& \qwbundle{n+1} & \gate{\text{$\mathcal{S}_0$}} & 
\end{quantikz}
= \quad
\begin{quantikz}
 & \gate{\text{X}} & \gate{\text{Z}} & \gate{\text{X}} & \\[-0cm]
 &  & \octrl{1}\wire[u]{q} &  &  \\[-0cm]
\setwiretype{n} & \vdots & \vdots & \vdots & \\[-0cm]
 &  & \octrl{-1} &  &
\end{quantikz}
\caption{Quantum circuit for the reflection operator. Here, X and Z denote the Pauli X and Pauli Z gates, respectively.}
\label{appB:fig2}
\end{figure}
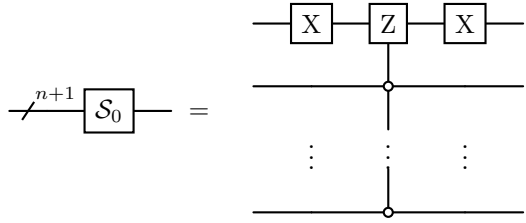

\begin{figure}
\centering
\begin{quantikz}
\lstick{$\ket{0}_{n+1}$} & \qwbundle{n+1} & \gate{\text{$\mathcal{A}_j[b_i]$}} & \gate{\text{$\mathcal{Q}_j^{k_i}[b_i]$}} & \meter{0/1}\arrow[r] & \rstick{$\ket{0}_{n+1}$} 
\end{quantikz}
%
%
\caption{Quantum circuit for the $i$-th iteration step in the RQAE algorithm \cite{Manzano.2023} for the $j$-th coefficients.}
\label{appB:fig3}
\end{figure}
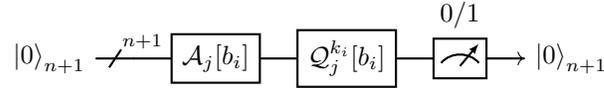

Quantum amplitude estimation is a QAA-based algorithm for estimating the amplitude of a desired state $\ket{\Phi_0}$. It achieves a quadratic speedup in the error bound compared to the direct sampling method, e.g., \cite{Manzano.2023}.
In this paper, we intend to apply the QAE algorithm to estimate the real-valued amplitude $\psi_j$ of each basis state $\ket{j}$, which is related to the value of the underlying function at each grid point. Then, it is sufficient to take $\ket{\Phi_0} = \ket{0}$ and consider a shifted oracle:
$$
U_{\psi_j} \ket{0}_n = \psi_j \ket{0}_n + \sqrt{1-|\psi_j|^2}\ket{0^\perp}_n, \quad j=0,\ldots,N-1. 
$$
Here, $\ket{0^\perp}$ denotes an arbitrary state that is perpendicular to the zero state. Assume that we are given the oracle preparing the quantum state $\ket{\psi}_n$:
$$
U_\psi \ket{0}_n = \ket{\psi}_n=\sum_{k=0}^{N-1}\psi_k \ket{k}_n.
$$
Applying an inverse of quantum modular adder (i.e., a quantum modular subtractor) \cite{Li.2020, Yuan.2023} defined by
$$
U_{\text{MADD}}^\dag[j] \ket{k}_n := \ket{k-j\ \mathrm{mod}\ N}_n,
$$
the shifted oracle can be obtained by $U_{\psi_j} = U_{\text{MADD}}^\dag[j] U_\psi$. 
For the QAE algorithm, we choose the well-performed algorithm RQAE proposed in \cite{Manzano.2023}. In order to determine the sign of the amplitude, we need to include a shift of the amplitude. To do this, we use an ancillary qubit and introduce the following oracle $\mathcal{A}_j[b]$: 
\begin{align*}
\mathcal{A}_j[b] \ket{0}_{n+1} 
&= \frac{\psi_j+b}{2}\ket{0}_{n+1} + \ket{0}\otimes \left(\frac{\sqrt{1-|\psi_j|^2}}{2}\ket{0^\perp}_n+\frac12 \sin(\theta_b/2)\ket{0}_{n-1}\otimes \ket{1}\right) \\
&\quad + \ket{1}\otimes \left(\frac{-\psi_j+b}{2}\ket{0}_n + \frac12 \sin(\theta_b/2)\ket{0}_{n-1}\otimes \ket{1}-\frac{\sqrt{1-|\psi_j|^2}}{2}\ket{0^\perp}_n\right)\\
&= \frac{\psi_j+b}{2}\ket{0}_{n+1} + \sqrt{1-\frac{(\psi_j+b)^2}{4}}\ket{0^\perp}_{n+1},
\end{align*}
which is realized by the quantum circuit in Fig.~\ref{appB:fig1}.
Here, $b\in [-1,1]$ is a parameter and $\theta_b:=2\arccos(b)\in [0,2\pi]$. In this case, $\ket{\Phi_0}$ is regarded as the zero state. Thus, the corresponding Grover operator can be defined by
$$
\mathcal{Q}_j[b] := -\mathcal{A}_j[b] \mathcal{S}_0 \mathcal{A}_j^\dag[b] \mathcal{S}_0, \quad \mathcal{S}_0 := I - 2\ket{0}\bra{0}.
$$
Here, the reflection operator $\mathcal{S}_0$ can be readily implemented using the quantum circuit in Fig.~\ref{appB:fig2}. 
In the $i$-th iteration step of the RQAE algorithm, we calculate the parameters $b_i\in [-1,1]$ and $k_i\in \mathbb{N}$. Then, the sampling result of the quantum circuit in Fig.~\ref{appB:fig3} helps us to estimate the $j$-th amplitude. One stops if the statistical error is sufficiently small. See \cite{Manzano.2023} for the details. 

\begin{figure}
\centering
\resizebox{15cm}{!}{
\begin{quantikz}
& \qwbundle{\!n_1+1} & \gate[3]{\text{$\mathcal{\tilde A}_j[b]$}} &  \\[-0cm]
& \qwbundle{\!n_2+1} &  &  \\[-0cm]
& \qw &  & 
\end{quantikz}
= \quad
\begin{quantikz}[transparent]
 &  & \gate[4]{U_\psi} & \gate[3]{U_{\text{ext}}} &  & \gate[3]{U_{\text{QFT}}^\dag} &  & \gate[5]{U_{\text{MADD}}^\dag[j]} & \gate{R_y(\theta_b)} &  & \\[-0cm]
 & \qwbundle{n_1-1} &  &  &  &  &  &  &  &  & \\[-0cm]
 & \qw & \linethrough &  &  &  &  &  &  &  & \\[-0cm]
 & \qwbundle{n_2} &  &  & \gate[2]{U_{\text{ext}}} &  & \gate[2]{U_{\text{QFT}}^\dag} &  &  &  & \\[-0cm]
 & \qw &  &  &  &  &  &  &  &  & \\[-0cm]
 & \gate{H} & \ctrl{-2} & \ctrl{-3} & \ctrl{-1} & \ctrl{-3} & \ctrl{-1} & \ctrl{-1} & \octrl{-5} & \gate{H} &
\end{quantikz}
}
\caption{Quantum circuit for the oracle $\mathcal{\tilde A}_j[b]$ for a 2D case. Here, $j\in \{0,\ldots,N-1\}$ and $b\in [-1,1]$ are given parameters, while $\theta_b:=2\arccos(b)$ is an angle calculated from $b$. }
\label{appB:fig4}
\end{figure}
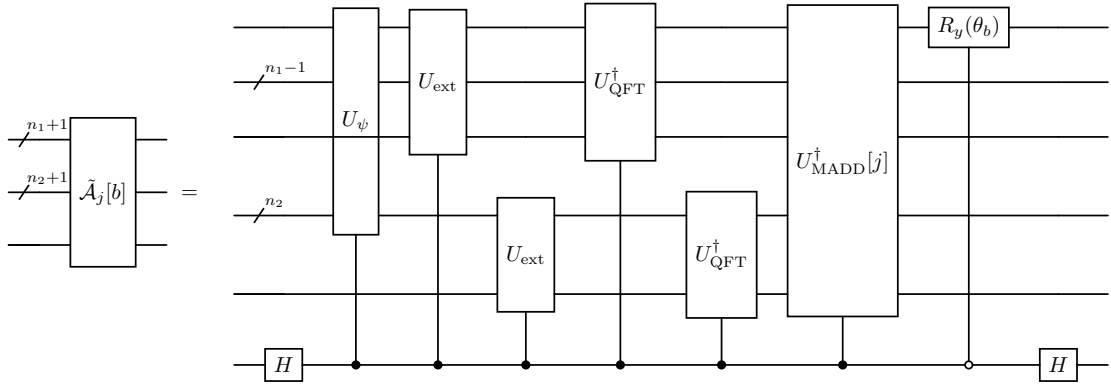

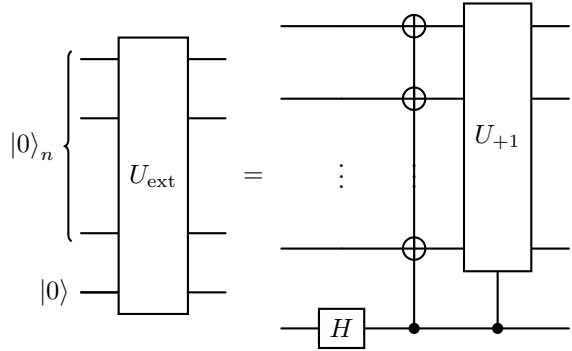
\begin{figure}
\centering
\begin{quantikz}
\lstick[4]{$\ket{0}_n$} &  \gate[5]{U_{\text{ext}}} & \\
&  & \\
\setwiretype{n} & \vdots &  \\
&  & \\
\lstick{\ket{0}} & \qw &  
\end{quantikz}
= 
\begin{quantikz}
&  & \targ{} & \gate[4]{U_{+1}} & \\[-0cm]
&  & \targ{} &  & \\[-0cm]
\setwiretype{n} & \vdots & \vdots &  & \\
&  & \targ{} &  & \\[-0cm]
& \gate{H} & \ctrl{-4} & \ctrl{-1} & 
\end{quantikz}
\caption{A quantum circuit for an even extension operator $U_{\text{ext}}$. Here, $U_{+1}$ denotes the quantum incrementer gate.}
\label{appB:fig5}
\end{figure}

On the other hand, the quantum circuits for obtaining the Fourier coefficient can be similarly constructed. The only difference is to substitute the oracle $\mathcal{A}_j[b]$ with an oracle $\mathcal{\tilde A}_j[b]$, see Fig.~\ref{appB:fig4}. 
Here, we apply an even extension operator $U_{\text{ext}}$ in Fig.~\ref{appB:fig5} so that the Fourier coefficients become real-valued \cite{Huang.2025pre}. 

\section{Theoretical details}
\label{sec:appC}

In this section, we discuss the theoretical upper bound of the quantum complexity for each method mentioned in Sect.~\ref{sec:2}. Although we investigate only the $l^2$ error for the normalized function (L2NS), the other errors, e.g., the maximum error and the root mean square error (RMSE), can be discussed similarly. 

Given a quantum state: $\sum_{k=0}^{N-1}\psi_k\ket{k}_n$ and a number of shots $N_{\text{shot}}$. The direct sampling method is to repeat the Z-basis measurements for all the qubits for $N_{\text{shot}}$ times. 
Let $p_k=|\psi_k|^2$ be the squared amplitude regarding the base state $\ket{k}$. The sampling error can be estimated as follows \cite{Huang.2025pre}:
$$
|\tilde p_k-p_k| \le \beta \sqrt{p_k(1-p_k)/N_{\text{shot}}},
$$
for some sufficiently large $\beta>0$. Here, $\tilde p_k$ is the estimated squared amplitude. As we mentioned, we use the natural correspondence between a number $k$ and a vector $\mathbf{k}=(k_1,\ldots,k_d)$ by $k=\sum_{\ell=1}^d k_\ell\left(\prod_{s=1}^{\ell-1} N_{s}\right)$ for $k=0,\ldots,N-1$, or $k=\sum_{\ell=1}^d k_\ell\left(\prod_{s=1}^{\ell-1} M_{s}\right)$ for $k=0,\ldots,M-1$

\noindent \underline{\bf RSR method}\ Assume $f\ge 0$. We have $\psi_k = f(\mathbf{x}_k)/A_N$, $p_k=|f(\mathbf{x}_k)|^2/A_N^2$ for $k=0,\ldots,N-1$ where $A_N:=\left(\sum_{k=0}^{N-1}|f(\mathbf{x}_k)|^2\right)^{1/2}$. Then, the normalized function is approximated as follows:
$$
f(\mathbf{x}_j)/A_N \approx \sqrt{\tilde p_j}, \quad j=0,\ldots,N-1.
$$
Employing the equality $|a-b| = \frac{|a^2-b^2|}{|a+b|}$ for $a,b\ge 0$, we estimate
\begin{align*}
\sum_{j=0}^{N-1} \left|\frac{f(\mathbf{x}_j)}{A_N}-\sqrt{\tilde p_j}\right|^2 
= \sum_{j=0}^{N-1} \frac{\left(|f(\mathbf{x}_j)|^2/A_N^2-\tilde p_j\right)^2}{|f(\mathbf{x}_j)/A_N+\sqrt{\tilde p_j}|^2} 
\le \sum_{j=0}^{N-1} \frac{\beta^2p_j(1-p_j)}{N_{\text{shot}}|f(\mathbf{x}_j)/A_N|^2}
= \frac{\beta^2 (N-1)}{N_\text{shot}}.
\end{align*}
Thus, the $l^2$ error for the normalized solution can be estimated:
\begin{align*}
& \text{Err}_{\text{L2NS}} = \left(\sum_{j=0}^{N-1}\left|\frac{f(x_j)}{A_N}-\sqrt{\tilde p_j}\right|^2\right)^{1/2} \le \frac{\beta \sqrt{N}}{\sqrt{N_{\text{shot}}}}.
\end{align*}
Then, the number of shots satisfies $N_{\text{shot}} = O\left(N(1/\varepsilon)^{2}\right)$ so as to achieve the error bound $\varepsilon$ in the L2NS. In each shot, we have one query to the oracle for preparing the quantum state. Therefore, the quantum complexity is $O\left(N(1/\varepsilon)^{2}\right)$. 

\noindent \underline{\bf FSR method}\ We have $\psi_k=c_{k}$ and $p_k=|c_k|^2$, which is the case of the original FSR method. For the modified FSR method in Appendix \ref{sec:appA}, we can discuss similarly by dividing the Fourier coefficients into their real and imaginary parts. According to Eq.~\eqref{sec2:eq-cfr}, the approximate function is given as follows:
$$
\frac{f(\mathbf{x}_j)}{A_N} \approx \tilde{f}_{j,M} := \frac{1}{\sqrt{N}}\sum_{k_1=-(M_1-1)}^{M_1-1}\cdots \sum_{k_d=-(M_d-1)}^{M_d-1} \mathrm{sgn}(c_{k})\sqrt{\tilde p_k} \prod_{\ell=1}^d \mathrm{exp}\left(\mathrm{i}\frac{2\pi}{N_\ell} k_\ell j_\ell\right), \quad j=0,\ldots,N-1.
$$
Here, $\tilde p_k := \tilde p_{-k}$ for $k=-(M-1),\ldots,-1$, and we assume that the signs are correctly determined for simplicity. 
Then, using the triangle inequality, we estimate
\begin{align}
\nonumber
&\quad \sum_{j=0}^{N-1} \left|\frac{f(x_j)}{A_N}-\tilde f_{j,M}\right|^2 \\
\nonumber
&\le \frac{2}{N} \sum_{j=0}^{N-1} \left|\sum_{s=1}^d\sum_{\mathbf{k}\in \mathcal{T}_s}c_k \prod_{\ell=1}^d\mathrm{exp}\left(\mathrm{i}\frac{2\pi}{N_\ell} k_\ell j_\ell\right)\right|^2 \\
\nonumber
&\quad + \frac{2}{N} \sum_{j=0}^{N-1} \left|\sum_{k_1=-(M_1-1)}^{M_1-1} \cdots \sum_{k_d=-(M_d-1)}^{M_d-1} \mathrm{sgn}(c_k)\left(\sqrt{p_k}-\sqrt{\tilde p_k}\right)\prod_{\ell=1}^d \mathrm{exp}\left(\mathrm{i}\frac{2\pi}{N_\ell} k_\ell j_\ell\right)\right|^2 \\
\nonumber
&= 2\sum_{s=1}^d\sum_{\mathbf{k}\in \mathcal{T}_s} |c_k|^2 + 2\sum_{k_1=-(M_1-1)}^{M_1-1} \cdots \sum_{k_d=-(M_d-1)}^{M_d-1} |\mathrm{sgn}(c_k)|^2\left|\sqrt{p_k}-\sqrt{\tilde p_k}\right|^2 \\
\label{appC:eq-ck}
&\le 2\sum_{s=1}^d\sum_{\mathbf{k}\in \mathcal{T}_s} |c_k|^2 + \frac{2^{d+1}\beta^2 M}{N_{\text{shot}}}.
\end{align}
Here, $M:=\prod_{\ell=1}^d M_\ell$ and we define the partitions:
$$
\mathcal{T}_s := \{(k_1,\ldots,k_d); k_t\in \mathcal{K}_t^{(1)} \text{ for any } t=1,\ldots,s-1,\ k_s\in \mathcal{K}_s^{(2)},\text{ and } k_t\in \mathcal{K}_t \text{ for any } t=s+1,\ldots,d\},
$$
for any $s=1,\ldots,d$. In the above definition, we use the notations:
\begin{align*}
&\mathcal{K}_\ell := \{-N_\ell/2,\ldots,N_\ell/2-1\}, \quad
\mathcal{K}_\ell^{(1)} := \{-M_\ell+1,\ldots,M_\ell-1\}, \quad \\
&\mathcal{K}_\ell^{(2)} := \{-N_\ell/2,-N_\ell/2+1\ldots,-M_\ell, M_\ell,M_\ell+1,\ldots,N_\ell/2-1\},
\end{align*}
for $\ell=1,\ldots,d$. Following the strategy of Lemma 1 in \cite{Huang.2025pre}, we can use the integration by parts to obtain the decay of the Fourier coefficients: $|c_k| = O(k_1^{-p} \cdots k_d^{-p})$ for some integer $p\ge 1$, provided that the function is sufficiently smooth and satisfies some periodic boundary conditions. Note that we should remove the decay $k_\ell^{-p}$ if $k_\ell=0$. Thus, we can estimate the squared Fourier coefficients in each $\mathcal{T}_s$: 
\begin{align*}
\sum_{\mathbf{k}\in \mathcal{T}_1} |c_k|^2 &\le 2^d \sum_{k_1=M_1}^{N_1/2} \sum_{k_2=0}^{N_2/2} \cdots \sum_{k_d=0}^{N_d/2} |c_k|^2 \le C_d \sum_{k_1=M_1}^{N_1/2} \sum_{k_2=0}^{N_2/2} \cdots \sum_{k_d=0}^{N_d/2} (\max\{k_1,1\})^{-2p} \cdots (\max\{k_d,1\})^{-2p}\\
&= C_d \left(M_1^{-2p}+\sum_{k_1=M_1+1}^{N_1/2} k_1^{-2p}\right) \prod_{\ell=2}^d \left(2+\sum_{k_\ell=2}^{N_\ell/2}k_\ell^{-2p}\right) \\
&\le C_d \left(M_1^{-2p}+\int_{M_1+1}^{+\infty} (x-1)^{-2p} \mathrm{d}x\right) \left(2+\int_{2}^{+\infty} (x-1)^{-2p} \mathrm{d}x\right)^{d-1} \\
&= C_{d} \left(M_1^{-2p}+\frac{M_1^{-2p+1}}{2p-1}\right)\left(\frac{4p-1}{2p-1}\right)^{d-1} \le C_{p,d} M_1^{-(2p-1)},
\end{align*}
for some positive constants $C_d, C_{p,d}$ possibly depending on $p$ and $d$. Similarly, we estimate
$$
\sum_{\mathbf{k}\in \mathcal{T}_s} |c_k|^2 \le C_{p,d,s} \prod_{t=1}^s M_t^{-(2p-1)} \le C_{p,d,s} M_1^{-(2p-1)}.
$$
Since we can change the spatial index $\ell$ to define a suitable partition, we assume that $M_1 = M_{\text{max}} := \max_{\ell=1,\ldots,d} M_\ell$ without loss of generality. 
Hence, we obtain the estimate for the L2NS: 
\begin{align*}
\left(\sum_{j=0}^{N-1} \left|\frac{f(\mathbf{x}_j)}{A_N}-\tilde f_{j,M}\right|^2\right)^{1/2} \le C\sqrt{M_{\text{max}}^{-(2p-1)}+\frac{M_{\text{max}}^d}{N_{\text{shot}}}}.
\end{align*}
We minimize the right-hand side by taking $M_{\text{max}}=O\left(N_{\text{shot}}^{1/(2p-1+d)}\right)$. 
This yields the estimation of the number of shots: $N_{\text{shot}}=O\left(\varepsilon^{-(2+2d/(2p-1))}\right)$, which is proportional to the quantum complexity. In other words, $M=O\left(\varepsilon^{-2d/(2p-1)}\right)$. We note that the (controlled) QFTs, the incrementer gates, as well as the modular adder, are both quantum efficient and have a complexity of at most $O((\log_2 N)^2)$. Compared to $\mathcal{C}_\psi$, it does not change the order of the total complexity in each quantum circuit. 
\begin{remark}[Improved order for the FSR method]
\label{appC:rem6}
The above discussion gives an upper bound on the required number of shots. In fact, the second term in the last inequality of Eq.~\eqref{appC:eq-ck} is overestimated. Since we use a sampling-based method, $\tilde p_k$ will be zero with probability $1-\delta$ for any $k$ such that $|c_k|^2\le \beta_\delta/N_{\mathrm{shot}}$ ($\beta_\delta\in (0,1)$ depends on $\delta$). For such $k$, the error is bound by $|c_k|^2=O(k_1^{-2p}\cdots k_d^{-2p})$ instead of $\beta^2/N_{\mathrm{shot}}$. 
Let $M_1=\cdots=M_d=M_0$ for simplicity. By the (adaptive) FSR method, we have $M_0=O(N_{\mathrm{shot}}^{1/2p})$. In this setting, we can estimate carefully the second term to obtain an upper bound $O\left(N_{\mathrm{shot}}^{-(2p-1)/2p}(\mathrm{ln}\, N_{\mathrm{shot}}^{1/2p})^{d-1}\right)$. This implies that 
\begin{align*}
\left(\sum_{j=0}^{N-1} \left|\frac{f(\mathbf{x}_j)}{A_N}-\tilde f_{j,M}\right|^2\right)^{1/2} \le CN_{\mathrm{shot}}^{-\frac{2p-1}{4p}} (\mathrm{ln}\, N_{\mathrm{shot}})^{\frac{d-1}{2}}.
\end{align*}
Recalling that the logarithmic increase is minor compared to the polynomial decay, the order of the FSR method can be improved to $N_{\mathrm{shot}}=\tilde O\left(\varepsilon^{-(2+2/(2p-1))}\right)$. 
\end{remark}

\noindent \underline{\bf ARSR method}\ Assume $f\ge 0$. For $\mathbf{\bar k}=(\bar k_1,\ldots,\bar k_d)$, $\bar k_\ell=0,\ldots,M_\ell-1$, $\ell=1,\ldots,d$, we let $\mathbf{k}=(k_1,\ldots,k_d)$ with $k_\ell = \bar{k}_\ell N_\ell/M_\ell$. We introduce the scaled mean square of the function as follows:
$$
g_{\mathbf{\bar k}} := \sum_{j_1=0}^{N_1/M_1-1}\cdots \sum_{j_d=0}^{N_d/M_d-1} \left|f\left((k_1+j_1)L_1/N_1,\ldots,(k_d+j_d)L_d/N_d\right)/A_N\right|^2.
$$
By the direct sampling method using the quantum circuit in Fig.~\ref{sec2:fig2}, we have $p_{\bar k}=g_{\mathbf{\bar k}}$. 
Moreover, $\sqrt{g_{\mathbf{\bar k}}M/N}$ gives the RMS approximation of the normalized function $f/A_N$ at the grid points $\mathbf{\hat x}_{\mathbf{\bar k}}=\left((\bar k_1+1/2-M_1/(2N_1))L_1/M_1,\ldots,(\bar k_d+1/2-M_d/(2N_d))L_d/M_d\right)$ (provided that $2\le M_\ell \le N_\ell/2$). Thus, we define $\tilde f_{\bar k,M}=\sqrt{\tilde p_{\bar k}M/N}$ we have
\begin{align*}
\left|\frac{f(\mathbf{\hat x}_{\mathbf{\bar k}})}{A_N}-\tilde f_{\bar k,M}\right| \le 
\left|\frac{f(\mathbf{\hat x}_{\mathbf{\bar k}})}{A_N}-\sqrt{\frac{M}{N} g_{\mathbf{\bar k}}}\right| + \sqrt{\frac{M}{N}}\left|\sqrt{\tilde p_{\bar k}}-\sqrt{g_{\mathbf{\bar k}}}\right| =: I_1+I_2.
\end{align*}
The first term is the RMS approximation error, while the second term is the sampling/statistic error that depends on the number of shots $N_{\text{shot}}$. Similar to the above discussions, the second term satisfies
$$
I_2 \le C\sqrt{\frac{M}{N}} \sqrt{\frac{1-g_{\mathbf{\bar k}}}{N_{\text{shot}}}}.
$$
On the other hand, if $|f(\mathbf{x})|\ge f_0>0$ which is independent of $M$, then we have
\begin{align*}
I_1 &\le \left|\frac{|f(\mathbf{\hat x}_{\mathbf{\bar k}})|^2}{|A_N|^2}-\frac{M}{N} g_{\mathbf{\bar k}}\right|/\left|\frac{f(\mathbf{\hat x}_{\mathbf{\bar k}})}{A_N}\right| \\
&\le \frac{1}{A_N f_0} \frac{M}{N} \left|\sum_{j_1=0}^{N_1/M_1-1}\cdots \sum_{j_d=0}^{N_d/M_d-1} \left|f(\mathbf{x}_{\mathbf{\bar k},\mathbf{j}})\right|^2 - |f(\mathbf{\hat x}_{\mathbf{\bar k}})|^2\right|. 
\end{align*}
Here, we denote $\mathbf{x}_{\mathbf{\bar k},\mathbf{j}} := \left((k_1+j_1)L_1/N_1,\ldots,(k_d+j_d)L_d/N_d\right)$. 
For sufficiently smooth $f$ (e.g., $f\in C^2$), we employ the Taylor expansion at $\mathbf{\hat x}_{\mathbf{\bar k}}$ for each $\mathbf{\bar k}$: 
\begin{align*}
f(\mathbf{x}_{\mathbf{\bar k},\mathbf{j}}) &= f(\mathbf{\hat x}_{\mathbf{\bar k}}) + \sum_{\ell=1}^d \partial_{x_\ell} f(\mathbf{\hat x}_{\mathbf{\bar k}}) \left(j_\ell+\frac{1}{2}-\frac{N_\ell}{2M_\ell}\right)\frac{L_\ell}{N_\ell} \\
&\quad + \frac{1}{2}\sum_{\ell,s=1}^d \partial_{x_{\ell}}\partial_{x_{s}} f(\mathbf{\xi}_{\mathbf{j}}) \left(j_\ell+\frac{1}{2}-\frac{N_\ell}{2M_\ell}\right)\left(j_s+\frac{1}{2}-\frac{N_s}{2M_s}\right)\frac{L_\ell L_s}{N_\ell N_s},
\end{align*}
where $\mathbf{\xi}_{\mathbf{j}}$ is some point that is close to $\mathbf{\hat x}_{\mathbf{\bar k}}$. For simplicity, we assume $L_1=\cdots=L_d=L_0$, $N_1=\cdots=N_d=N_0$, and $M_1=\cdots=M_d=M_0$. 
With a direct calculation, we have 
$$
I_1 \le \frac{C_{f,d}}{A_N M_0^2}.
$$
Otherwise, if the function $f$ has zeros, then there are some points $\mathbf{\hat x}_{\mathbf{\bar k}}$ such that $f(\mathbf{\hat x}_{\mathbf{\bar k}}) \le O(1/M_0)$. Hence, a naive calculation yields $I_1\le C/(A_N M_0)$. 
Whereas, if we have additional assumption that $\partial_{\mathbf{x}} f(\mathbf{x})=0$ for $f(\mathbf{x})=0$, then the first and second terms in the Taylor expansion vanishes, and we still obtain the second-order decay: $I_1 \le \frac{C}{A_N M_0^2}$.  
To approximate $f(\mathbf{x}_j)/A_N$, we can use spline interpolation (e.g., linear interpolation). The approximate normalized function is denoted by $\hat f_{j}$ for $j=0,\ldots,N-1$. 
Recalling that the spline method leads to at least a quadratic decay with respect to the diameter of the sub-interval (e.g., \cite{Agarwal.1991}), we have   
\begin{align*}
\left|\hat f_j-\frac{f(\mathbf{x}_j)}{A_N}\right| 
&\le \left|\frac{f_{\text{sp}}(\mathbf{x}_j)}{A_N}-\frac{f(\mathbf{x}_j)}{A_N}\right| + \left|\hat f_j-\frac{f_{\text{sp}}(\mathbf{x}_j)}{A_N}\right| \\
&\le \frac{C_f}{A_N M_0^2} + C\max_{\mathbf{\bar k}} \left|\frac{f(\mathbf{\hat x}_{\mathbf{\bar k}})}{A_N}-\tilde f_{\bar k,M}\right| \\
&\le \frac{C}{A_N M_0^2} + C\sqrt{\frac{M}{N N_{\text{shot}}}},
\end{align*}
for any $j=0,\ldots,N-1$. Here, $f_{\text{sp}}$ denotes the (linear) spline interpolation of the values of the function $f$ at the grid points $\{\mathbf{\hat x}_{\mathbf{\bar k}}\}_{\mathbf{\bar k}}$. 
Therefore, we estimate the L2NS as follows:
\begin{align*}
\left(\sum_{j=0}^{N-1} \left|\hat f_j-\frac{f(\mathbf{x}_j)}{A_N}\right|^2\right)^{1/2} &\le \left(N \max_{j=0,\ldots,N-1} \left|\hat f_j-\frac{f(\mathbf{x}_j)}{A_N}\right|^2\right)^{1/2} \\
&\le C\left(\frac{1}{M_0^4} + \frac{M_0^d}{N_{\text{shot}}}\right)^{1/2}. 
\end{align*}
We choose $M_0 = O\left(N_{\text{shot}}^{1/(d+4)}\right)$ and obtain the overhead $N_{\text{shot}} = O(\varepsilon^{-(2+d/2)})$. In this case, the approximation parameter satisfies $M=M_0^d=O(\varepsilon^{-d/2})$.

\noindent \underline{\bf RSQAE method}\ We employ the RQAE algorithm in \cite{Manzano.2023} since it also determines the sign of the real-valued amplitude. The complexity analysis was provided in Theorem 3.1 therein. Here, we discuss the total complexity of the application to our problem. For each amplitude $\psi_j$, $j=0,\ldots,N-1$, we use the RQAE algorithm once. By choosing the amplification policy $q=2$ (i.e., the amplification ratio of $k$ in the iterations for each RQAE) and the error bound $\epsilon_0$: $|\tilde\psi_j-\psi_j|\le \epsilon_0$, the maximal circuit depth is $O(1/\epsilon_0)$, and the total number of queries to the oracle $U_{\psi_j}$ is $O\left((1/\epsilon_0)\log(\log(1/\epsilon_0))\right)=\tilde{O}\left(1/\epsilon_0\right)$, where the double logarithmic dependence comes from the required number of shots in each iteration. For both the L2NS and the maximum error, we have the scaling factor $\sqrt{N}$: 
\begin{align*}
& \text{Err}_{\text{L2NS}} = \left(\sum_{j=0}^{N-1}\left|\frac{f(x_j)}{A_N}-\tilde \psi_j\right|^2\right)^{1/2} \le \sqrt{N\epsilon_0^2}, 
\quad \text{Err}_{\text{max}} = \max_{j\in \mathcal{J}} \left|f(x_j)-A_N\tilde \psi_j\right| \le \sqrt{N}\epsilon_0. 
\end{align*}
This implies $\epsilon_0=O\left(\varepsilon/\sqrt{N}\right)$, and we obtain the quantum complexity $\tilde{O}(\sqrt{N}/\varepsilon)$ for each amplitude. In the case that there are $J$ target grid points, we conclude the total quantum complexity $\tilde{O}\left(J\sqrt{N}/\varepsilon\right)$. 
On the other hand, the classical complexity of making the histogram is proportional to the number of target points $J$, the number of iterations $O\left(\log_2 (1/\epsilon_0)\right)$, and the number of shots in each iteration $O\left(\log(\log(1/\epsilon_0))\right)$, which is $O\left(J \mathrm{polylog}\left(\sqrt{N}/\varepsilon\right)\right)$.  

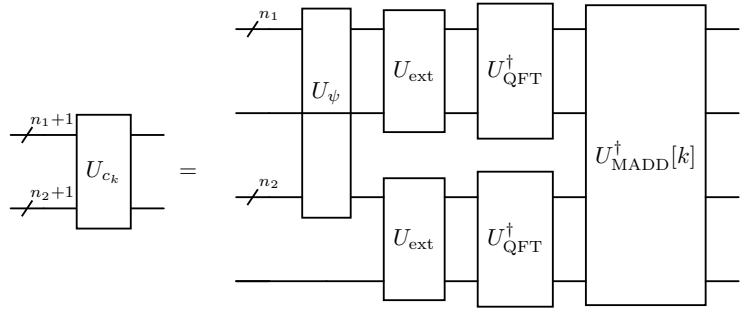
\begin{figure}
\centering
\resizebox{10cm}{!}{
\begin{quantikz}
& \qwbundle{\!n_1+1} & \gate[2]{\text{$U_{c_k}$}} &  \\[-0cm]
& \qwbundle{\!n_2+1} &  & 
\end{quantikz}
= \quad
\begin{quantikz}[transparent]
 & \qwbundle{n_1} & \gate[3, label
style={yshift=0.3cm}]{U_\psi} & \gate[2]{U_{\text{ext}}} & \gate[2]{U_{\text{QFT}}^\dag} & \gate[4]{U_{\text{MADD}}^\dag[k]} & \\[-0cm]
 &  & \linethrough &  &  &  & \\[-0cm]
 & \qwbundle{n_2} &  & \gate[2]{U_{\text{ext}}} & \gate[2]{U_{\text{QFT}}^\dag} &  & \\[-0cm]
 & \qw &  &  &  &  & \\[-0cm]
\end{quantikz}
}
\caption{Quantum circuit for the oracle $U_{c_k}$ for a 2D case. The extension operator (Fig.~\ref{appB:fig5}) is added so that the Fourier coefficients of the extended function are real-valued.}
\label{appC:fig1}
\end{figure}

\noindent \underline{\bf FSQAE method}\ We use the RQAE method to estimate the Fourier coefficients. The difference from the RSQAE method is to change the oracle $U_{\psi_j}$ into an oracle $U_{c_k}$ (Fig.~\ref{appC:fig1}) that prepares the Fourier coefficients and shifts the $k$-th Fourier coefficient to the amplitude of the base state $\ket{0}$. We use the extension operator $U_{\text{ext}}$ (Fig.~\ref{appB:fig5}) proposed in \cite{Huang.2025pre} so that the Fourier coefficients for the extended function are real-valued. 
The gate complexity of the extension operators, the QFTs, and the modular adder is $O\left(n^2\right)$ where $n=\sum_{\ell=1}^d n_\ell$ is the total number of qubits for the input quantum state $\ket{\psi}$. Thus, the additional gates do not change the quantum complexity of the original oracle $U_\psi$. 
For the approximation parameters $M_1,\ldots,M_d$, we calculate the Fourier coefficients $c_k$ for all $k_\ell=0,\ldots,M_\ell-1$, $\ell=1,\ldots,d$. Employing the RQAE algorithm, we estimate each Fourier coefficient up to the error bound $\epsilon_0$ under the complexity $\tilde{O}(1/\epsilon_0)$. Similar to the FSR method, we denote the approximation to the normalized function by 
$$
\tilde{f}_{j,M} := \frac{1}{\sqrt{N}}\sum_{k_1=-(M_1-1)}^{M_1-1}\cdots \sum_{k_d=-(M_d-1)}^{M_d-1} \tilde c_k \prod_{\ell=1}^d \mathrm{exp}\left(\mathrm{i}\frac{2\pi}{N_\ell} k_\ell j_\ell\right), \quad j=0,\ldots,N-1,
$$
where $\tilde c_k$ is the Fourier coefficient derived by the RQAE algorithm. Thus, the L2NS is estimated as follows:
\begin{align*}
\mathrm{Err}_{\text{L2NS}} &= \left(\sum_{j=0}^{N-1} \left|\frac{f(x_j)}{A_N}-\tilde f_{j,M}\right|^2\right)^{1/2} \\
&\le \Bigg(\frac{2}{N} \sum_{j=0}^{N-1} \left|\sum_{s=1}^d\sum_{\mathbf{k}\in \mathcal{T}_s}c_k \prod_{\ell=1}^d\mathrm{exp}\left(\mathrm{i}\frac{2\pi}{N_\ell} k_\ell j_\ell\right)\right|^2 \\
&\quad + \frac{2}{N} \sum_{j=0}^{N-1} \left|\sum_{k_1=-(M_1-1)}^{M_1-1} \cdots \sum_{k_d=-(M_d-1)}^{M_d-1} \left(c_k-\tilde c_k\right)\prod_{\ell=1}^d \mathrm{exp}\left(\mathrm{i}\frac{2\pi}{N_\ell} k_\ell j_\ell\right)\right|^2\Bigg)^{1/2} \\
&= \left(2\sum_{s=1}^d\sum_{\mathbf{k}\in \mathcal{T}_s} |c_k|^2 + 2\sum_{k_1=-(M_1-1)}^{M_1-1} \cdots \sum_{k_d=-(M_d-1)}^{M_d-1}\left|c_k-\tilde c_k\right|^2\right)^{1/2} \\
&\le C\sqrt{M_{\text{max}}^{-(2p-1)} + M_{\text{max}}^d \epsilon_0^2}.
\end{align*}
Here, $M_{\text{max}}$ denotes the maximum of $M_\ell$. We choose $M_{\text{max}}=O\left((1/\epsilon_0)^{2/(d+2p-1)}\right)$ and obtain $\epsilon_0=\varepsilon^{1+d/(2p-1)}$. In other word, we have $M_{\text{max}}=O\left((1/\varepsilon)^{s}\right)$ for $s=2/(2p-1)$. Hence, we obtain the quantum complexity $\tilde{O}\left((1/\varepsilon)^{1+ds/2}\right)$ for each Fourier coefficient. Since we need $M\le M_{\text{max}}^d$ Fourier coefficients, the total quantum complexity for the FSQAE method is $\tilde{O}\left((1/\varepsilon)^{1+3ds/2}\right)$. 
On the other hand, the classical complexity of making the histogram is proportional to the number of coefficients $O\left((1/\varepsilon)^{ds}\right)$, the number of iterations $O\left(\log_2 (1/\epsilon_0)\right)$, and the number of shots in each iteration $O\left(\log(\log(1/\epsilon_0))\right)$, which is $\tilde{O}\left((1/\varepsilon)^{ds}\right)$. 
Different from the FSR method, the dependence on the dimension in the power of $1/\varepsilon$ is necessary because the dominant coefficients cannot be adaptively determined by the number of shots. One possibility to release the dependence, to some extent, is the cutoff of the minor coefficients similar to \cite[Algorithm 4]{Patterson.2025}. 

\section{Numerical details}
\label{sec:appC2}

\subsection{Numerical comparison of post-processing methods}
\label{subsec:C2-1}

\begin{figure}[htp]
\centering
\resizebox{13cm}{!}{
\includegraphics[keepaspectratio]{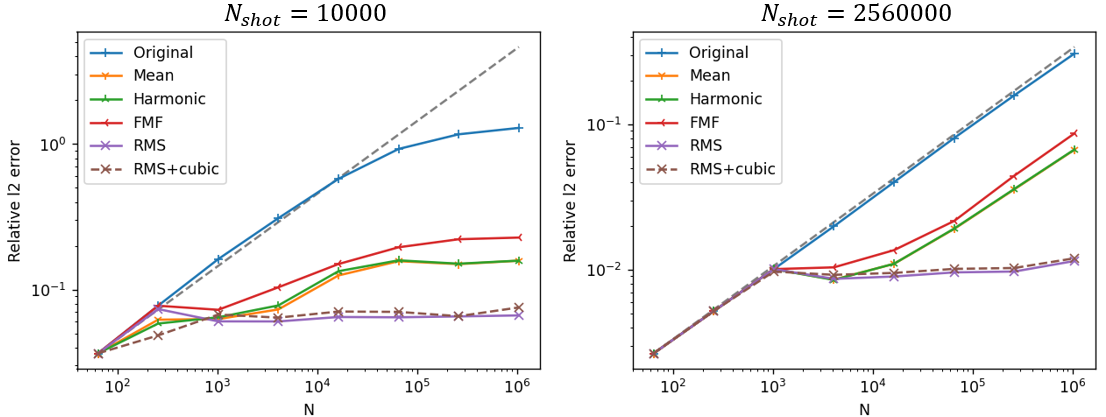}
}
\caption{Error plots for different post-processing methods in the example of a linear combination of 2D Gaussian functions. The left and right subplots show the results with $N_{\text{shot}}=10000$ and $N_{\text{shot}}=2560000$, respectively. The gray dashed lines denote the orders $O(\sqrt{N})$. }
\label{appC2:fig3}
\end{figure}

\begin{figure}[htp]
\centering
\resizebox{13cm}{!}{
\includegraphics[keepaspectratio]{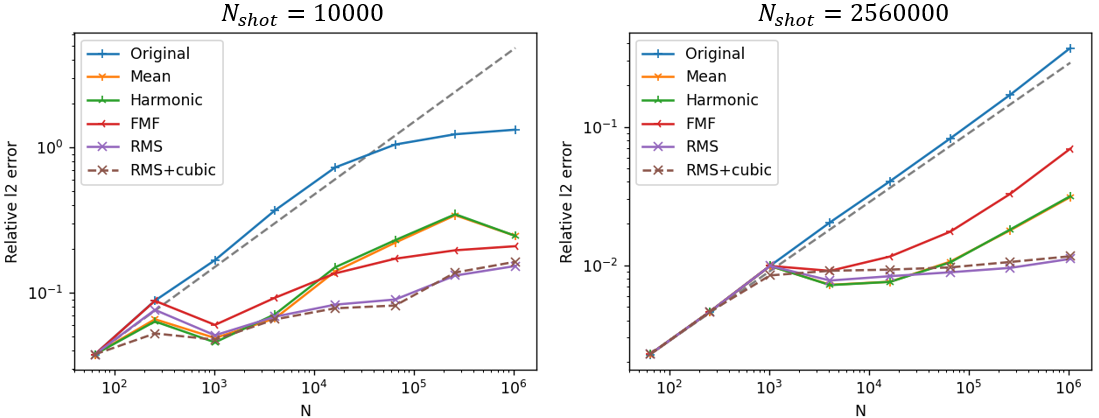}
}
\caption{Error plots for different post-processing methods in the example of a 2D sine function. The left and right subplots show the results with $N_{\text{shot}}=10000$ and $N_{\text{shot}}=2560000$, respectively. The gray dashed lines denote the orders $O(\sqrt{N})$. }
\label{appC2:fig4}
\end{figure}

In Sect.~\ref{subsec:2-2}, we mentioned that the proposed ARSR method is equivalent to the post-processing of the RSR results using the specific root mean square (RMS) approximation. Here, we assume the function is non-negative and consider the following averages for the post-processing:
\begin{itemize}
\item RMS: $\left(\frac{1}{N_a}\sum_{j=0}^{N_a-1} |a_j|^2\right)^{1/2}$;

\item Arithmetic average (Mean): $\frac{1}{N_a}\sum_{j=0}^{N_a-1} a_j$; 

\item Shifted harmonic average: $N_a/\left(\sum_{j=0}^{N_a-1} (a_j+\delta)^{-1}\right) -\delta$ with $\delta=0.1$ (shift parameter $\delta$ is introduced to avoid the singularity when $a_j=0$);

\item Fourth-root mean fourth-power (FMF) average: $\left(\frac{1}{N_a}\sum_{j=0}^{N_a-1} |a_j|^4\right)^{1/4}$.
\end{itemize}
Using the notation of the ARSR methods in Appendix \ref{sec:appC}, for each $\mathbf{\bar k}=(\bar k_1,\ldots,\bar k_d)$, $\bar k_\ell=0,\ldots,M_\ell-1$, $\ell=1,\ldots,d$, we note that $N_a = \prod_{\ell}^d (N_\ell/M_\ell)$ and 
$$
a_j^{(\mathbf{\bar k})} \leftrightarrow a_{j_1,\ldots,j_d}^{(\mathbf{\bar k})}= f\left(\bar k_1 L_1/M_1 + j_1 L_1/N_1,\ldots, \bar k_d L_d/M_d + j_d L_d/N_d\right),
$$
where $j\in \{0,1,\ldots,N_a-1\}$ has an one-to-one correspondence to the vector $\mathbf{j}=(j_1,\ldots,j_d)$, $j_\ell\in \{0,1,\ldots,N_\ell/M_\ell-1\}$ by the relation: 
$$
j = \sum_{\ell=1}^d j_\ell \left(\prod_{\ell^\prime=1}^{\ell-1}\frac{N_{\ell^\prime}}{M_{\ell^\prime}}\right).
$$
The change of the L2NS errors regarding the (total) grid number $N$ is demonstrated in Fig.~\ref{appC2:fig3} with an example of a linear combination of 2D Gaussian functions (Example 1 in Sect.~\ref{subsec:3-2}). 
Here, we set $N_1=N_2=N_0\in \{2^3, 2^4,\ldots, 2^{10}\}$ and $M_1=M_2=M_0$. To show the essential error, we choose $M_0\in \{2^1, 2^2, \ldots N_0\}$ for each post-processing method such that the L2NS error between the true and the RSR (with post-processing) solutions is minimized. We mainly employ linear interpolation in the post-processing, while we also include the cubic spline for the RMS post-processing (equivalent to the ARSR method). 
For a relatively small number of shots $N_\text{shot}=10000$, the orders of the post-processing methods are not clarified because the L2NS errors of the original RSR method are close to the upper bound ($l^2$ error between two normalized random vectors is usually bounded by $1$). Still, the RMS processing methods are obviously better than the others for sufficiently large $N$. For a relatively large number of shots $N_\text{shot}=2560000$, the asymptotic order $O(\sqrt{N})$ in the errors for all the post-processing methods except the RMS post-processing is confirmed. On the other hand, whatever spline interpolations are employed, the error for the RMS post-processing method is almost uniform regarding $N$. 
A similar plot for a 2D sine function (Example 2 in Sect.~\ref{subsec:3-2}) is shown in Fig.~\ref{appC2:fig4}. Although the error for the RMS post-processing seems to increase when $N_{\text{shot}}$ is small, it appears uniform for a relatively large number of shots. 

In the original RSR method, we use the sampling-based method to estimate the square of the amplitude. Then, the statistical error appears in the squared value of the (normalized) function. Therefore, only the RMS post-processing methods help to relieve the statistical errors and achieve $N$-uniform errors. Although the other post-processing methods reduce the error of the original RSR method to some extent, we conclude that only the RMS post-processing methods are efficient for large grid numbers, especially when we intend to achieve small errors.  

\subsection{Numerical comparison of methods for CFD solutions}
\label{subsec:C2-2}

\begin{figure}[htp]
\centering
\resizebox{13cm}{!}{
\includegraphics[keepaspectratio]{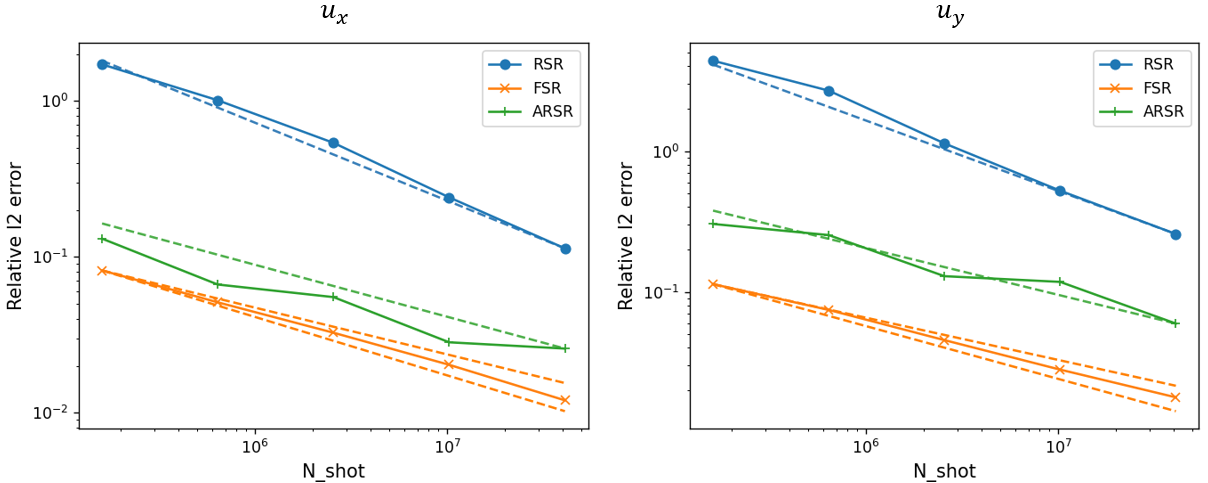}
}
\caption{Error plots for different methods in the example of a solution to a planar jet flow. The left and right subplots show the results for $u_x$ and $u_y$, respectively. The dashed lines denote the orders $O\left(1/N_{\text{shot}}^t\right)$ with $t=1/2,1/3,3/10,3/8$. }
\label{appC2:fig1}
\end{figure}

\begin{figure}[htp]
\centering
\resizebox{13cm}{!}{
\includegraphics[keepaspectratio]{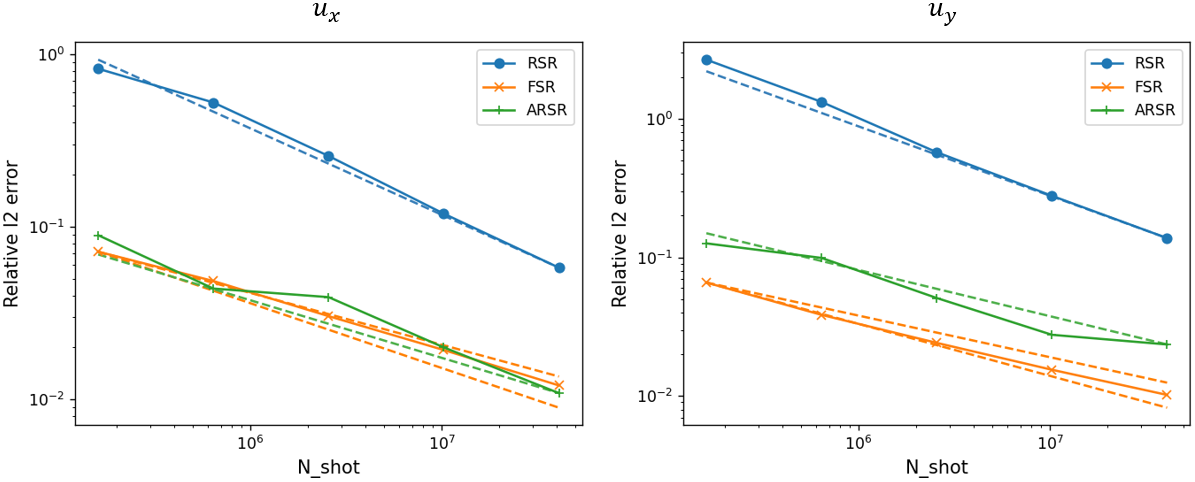}
}
\caption{Error plots for different methods in the example of a solution to a lid-driven cavity flow. The left and right subplots show the results for $u_x$ and $u_y$, respectively. The dashed lines denote the orders $O\left(1/N_{\text{shot}}^t\right)$ with $t=1/2,1/3,3/10,3/8$. }
\label{appC2:fig2}
\end{figure}

In Sect.~\ref{subsec:4-1}, we visualize the solutions to a planar jet flow and a lid-driven cavity flow with $N_{\text{shot}}=1.6\times 10^5,\, 4.096\times 10^7$. 
Here, we illustrate the decrease of the relative $l^2$ errors between the ``true" solution and the reconstructed solutions by different readout methods. The numerical decay orders for the jet flow and the cavity flow are shown in Fig.~\ref{appC2:fig1} and Fig.~\ref{appC2:fig2}, respectively. 
The numerical orders well fit the theoretical orders (dashed lines) in Table \ref{sec3:tab1}, and it indicates that the parameter $s=2/3$ for both CFD solutions. 
For the FSR method, we draw two dashed lines, where the upper line denotes the order $O\left(N_{\text{shot}}^{-(2p-1)/(4p-2+2d)}\right)$ by a rough estimation and the lower line denotes the order $O\left(N_{\text{shot}}^{-(2p-1)/(4p)}\right)$ by a careful calculation according to Remark \ref{appC:rem6}. Here, $p=2$ and thus $s=2/(2p-1)=2/3$. Although the asymptotic order ($N_{\text{shot}}\to \infty$) is the lower line, the presence of a logarithmic factor makes the numerical order behave between the above two lines for a moderate number of shots. 

\subsection{QAE-based Fourier readout method without extension}
\label{subsec:C2-3}

\begin{figure}
\centering
\resizebox{8cm}{!}{
\includegraphics[keepaspectratio]{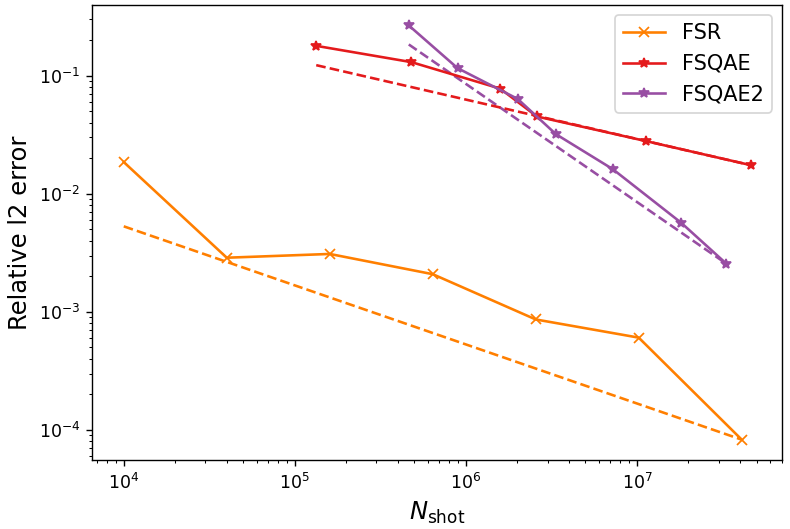}
}
\caption{Error plots for Fourier space readout methods in the example of a 2D trigonometric function. The number of shots is $N_{\text{shot}}\in 10000\times \{4^0,4^1,\ldots,4^6\}$. For the FSQAE and the FSQAE2 methods, the $x$-axis denotes the number of queries to the oracle, and the plot is done by taking $\epsilon_0\in \{0.05, 0.02, 0.01, 0.005, 0.0025, 0.001\}$ in the RQAE algorithm. 
The dashed lines denote the orders $O\left(1/N_{\text{shot}}^{t}\right)$ with $t=1/2,1/3, 1$ for the FSR, the FSQAE, and the FSQAE2 methods, respectively. }
\label{appC2:fig5}
\end{figure}

We provide an additional error plot for Example 2 in Sect.~\ref{subsec:3-2}. More precisely, we consider the QAE-based method regarding the modified FSR method (Appendix \ref{sec:appA}), that is, the QAE-based Fourier readout method without extension, and denote it by the FSQAE2 method. The error plot is shown in Fig.~\ref{appC2:fig5}. 
Since both the real and the imaginary parts of the Fourier coefficients are evaluated by the QAE algorithm in the FSQAE2 method, it has a larger pre-factor than the FSQAE method. However, for the sine function which is smooth and periodic. The even extension used in the FSQAE method reduces the regularity at the extended boundary, so the decay of the Fourier coefficients for the extended sine function is limited ($s=2/3$). In such a case, the FSQAE2 method demonstrates a better decay order with respect to the number of queries to the oracle, that is, $O(1/N_{\text{shot}})$. This decay order is even better than the FSR method in the orange line, and it implies that the QAE-based method can achieve the best possible quantum complexity $O(1/\varepsilon)$ provided that the function is smooth and periodic. 
Moreover, we mention that the pre-factors of the FSQAEs can be improved using the cutoff of the minor coefficients similar to \cite[Algorithm 4]{Patterson.2025}.

\section{Comments on time stepwise readout (TSR) strategy}
\label{sec:appD}

\subsection{Circuit implementation and complexity for TSR strategy}
\label{subsec:D-1}

We mention the circuit implementation of the strategy in Sect.~\ref{subsec:4-2}. It consists of three parts: the preparation of quantum states, the PITE operator, and the readout. For the first part, one can use the Fourier series loader (FSL) method proposed in \cite{Moosa.2023}, and the gate complexity is $O(\mathrm{polylog} N)$. Note that the FSL method fits the FSR method well because the FSR method provides exactly the approximate Fourier coefficients that are needed for the FSL method. For the second part, using the AAPITE algorithm \cite[Section 6]{Huang.2024pre}, it is sufficient to implement specific diagonal unitary matrices whose diagonal components depend on the approximate solutions. As we have obtained the dominant Fourier coefficients of the approximate solutions, it is possible to calculate the values at coarser grid points and construct the approximate diagonal unitary operator by the spline methods (e.g., \cite{Huang.2024}). The gate complexity remains $O(\mathrm{polylog} N)$ at the cost of introducing the PITE approximation error and the function approximation error. In the third part, the additional gate complexity is $O(\mathrm{polylog} N)$ and the number of shots has a polynomial dependence on the error bound. 
Noting that $K=T/\Delta t$ also has a polynomial dependence on the error bound, the proposed strategy above is efficient for a large grid number $N$, while it has a relatively large polynomial degree on the reciprocal of the error bound. 

On the other hand, it is natural to consider another time stepwise strategy where the second step is replaced by the following one:

\noindent $2^\prime)$\quad In the $k$-th time step for $k\ge 2$, we prepare the input quantum state corresponding to the initial conditions and apply the PITE operators for $(k-1)$ time steps where the nonlinear terms are implemented using the approximate solutions $\mathbf{\tilde u}^{(j)}$ from $j=0$ to $j=k-2$ ($\mathbf{\tilde u}^{(0)}:=\mathbf{u}^{(0)}$). Again, we post-select the ancillary qubit for the PITE algorithm to be $\ket{0}$ and employ the FSR method to reconstruct the approximate solutions at $t=k\Delta t$: $\mathbf{\tilde u}^{(k)}$.

Since Burgers' equation is a dissipative system whose solution has a decreasing norm, the success probability is strictly smaller than one. Moreover, by the quantum algorithm for Burgers' equation in \cite{Huang.2024pre}, one needs to introduce some scaling parameters so that the non-unitary operations can be implemented. This yields a constant loss of the success probability in each PITE step, and hence, the long-time simulation has an extremely small success probability. For example, the success probability at $T=1$ with $\Delta t=0.04$ is smaller than $1\times 10^{-7}$ ($\approx 5.3\times 10^{-8}$) in the case of Sect.~\ref{subsec:4-2}. 
The proposed strategy in Sect.~\ref{subsec:4-2} avoids the problem of decreasing success probability because the success probability has a lower bound larger than $0.3$ in each time step and greatly outperforms the one with the step $2^\prime)$.  

\subsection{Simulation details for Burgers' equation}
\label{subsec:D-2}

In Sect.~\ref{subsec:4-2}, we solved a 2D Burgers' equation by the proposed TSR strategy. 
We consider a grid number $N=2^5\times 2^5=1024$ and a small time interval $\Delta t=0.04$ so that the discretization error and the approximation error (including the Suzuki-Trotter error) are relatively small. Let $T=1$ and $K=T/\Delta t=25$. We calculate the reference solution at the $k$-th time step by solving the linearized Burgers' equation:
$$
\partial_t \mathbf{u}^{(k)} = \nu \nabla^2 \mathbf{u}^{(k)} - \mathbf{u}^{(k)}\cdot\nabla \mathbf{u}^{(k-1)}, 
$$
based on the matrix multiplications: 
$$
\mathbf{u}^{(k)} = \mathrm{exp}\left(-\Delta t H_{2N}^{(k-1)}\right) \mathbf{u}^{(k-1)}, \quad k=1,\ldots, K.
$$
Here, $H_{2N}^{(k-1)}\in \mathbb{R}^{2N\times 2N}$ is the discretization matrix satisfying 
$$
H_{2N}^{(k-1)} = 
\begin{pmatrix}
F_N D_N F_N^\dag & 0\\
0 & F_N D_N F_N^\dag
\end{pmatrix}
+
\begin{pmatrix}
P_{11}^{(k-1)} & P_{12}^{(k-1)} \\
P_{21}^{(k-1)} & P_{22}^{(k-1)}
\end{pmatrix},
$$ 
where $F_N$ is the matrix for the QFT, $D_N$ is the diagonal matrix for the diffusion part \cite{Huang.2024pre}, and $P_{ij}^{(k-1)}$ are diagonal matrices whose diagonal components are $\partial_x u_x^{(k-1)}, \partial_y u_x^{(k-1)}, \partial_x u_y^{(k-1)}$, and $\partial_y u_y^{(k-1)}$, respectively. The derivatives are calculated using the fast Fourier transform. Therefore, the reference solution does not include any errors in the quantum algorithms. 

On the other hand, for the quantum solution, we employ the approximate PITE algorithm proposed in \cite{Huang.2024pre}. For the diffusion part, we choose the cut-off parameter $64$ \cite[Appendix C]{Huang.2024pre}) so that the gate complexity is $O(\max\{n_1,n_2\}^3)$. For the potential part, we use the decomposition of the block diagonal matrices \cite[Section 6]{Huang.2024pre} and the precise implementation of the diagonal matrices \cite{Zhang.2024}. The gate complexity of the precise implementation is $O(N)$, but it is possible to use modified spline interpolation methods \cite{Huang.2024} to reduce the gate complexity to $O(\mathrm{polylog}N)$ at the expense of the approximation error. 
For the readout part, we apply the modified FSR method with the quantum circuits in Appendix \ref{sec:appA}. 

More precisely, we use the Qiskit simulator to get the precise statevector and the success probability after each time step and employ the modified FSR method to reconstruct the statevector for each direction using $1\times 10^5$ shots for both quantum circuits in Fig.~\ref{appA:fig3} and Fig.~\ref{appA:fig4}. Besides, taking the success probability $p_k$ of the PITE algorithm in the $k$-th time step into account, the number of shots in the $k$-th step is $(4\times 10^5)/p_k$. Since $0.30\le p_k \le 0.85$, we actually need about $1\times 10^6$ shots in each time step and totally about $2\times 10^7$ shots to derive the simulation results until $t=1$ in Fig.~\ref{sec4:fig5}. 
We remark that the final success probability at $t=1$ is smaller than $1\times 10^{-7}$ if we do not use the TSR strategy. Although the error will be slightly improved, the required number of shots is larger than $1\times 10^{12}$, which is huge compared to the TSR strategy. 
Although even the TSR strategy requires more than 10 million shots for the simulation, the quantum advantage lies in the $N$-independent scaling as the grid number $N$ increases to a sufficiently large number. 

\section{Required numbers of shots in detailed settings}
\label{sec:appF}

We compare the number of shots of the RSR, the ARSR, and the FSR methods for detailed dimensions and error bounds using the theoretical bounds in Table \ref{sec3:tab1}. 
Conventionally, the grid number can be transformed to the error bound by $N=O(1/\varepsilon^d)$ in the case of a first-order precision algorithm. 
\begin{table}[htb]
\centering
\caption{Detailed resource estimations for a piecewise $W^{1,1}$ function. }
\label{appF:tab1}
\scalebox{0.8}[0.8]{
\begin{tabular}{cccc|l}
\hline
Dimension & Error bound & Method & $^\sharp$Shots$^{[1]}$ & Acceleration$^{[1]}$ \\
\hline
$d=1$ & $\varepsilon=0.01$ & RSR/ARSR & 1e6 & $\times$ 1\\
&  & FSR & 1e8 & $\times$ 1/100\\
\cline{2-5}
& $\varepsilon=0.001$ & RSR/ARSR & 1e9 & $\times$ 1\\
&  & FSR & 1e12 & $\times$ 1/1000\\
\hline
$d=2$ & $\varepsilon=0.01$ & RSR/ARSR & 1e8 & $\times$ 1\\
&  & FSR & 4.6e8 & $\times$ 1/5\\
\cline{2-5}
& $\varepsilon=0.001$ & RSR/ARSR & 1e12 & $\times$ 1\\
&  & FSR & 6.9e12 & $\times$ 1/10\\
\hline
$d=3$ & $\varepsilon=0.01$ & RSR/ARSR & 1e10 & $\times$ 1\\
&  & FSR & 2.1e9 & $\times$ 5\\
\cline{2-5}
& $\varepsilon=0.001$ & RSR/ARSR & 1e15 & $\times$ 1\\
&  & FSR & 4.8e13 & $\times$ 21\\
\hline
\end{tabular}
}
\begin{flushleft}
\footnotesize [1] These columns only indicate the rough numbers of shots and the acceleration rates because the pre-factors depend on the properties (e.g., function norms) of the underlying function. 
\end{flushleft}
\end{table}
\begin{table}[ht]
\centering
\caption{Detailed resource estimations for a continuous piecewise $W^{2,1}$ function. }
\label{appF:tab2}
\scalebox{0.8}[0.8]{
\begin{tabular}{cccc|l}
\hline
Dimension & Error bound & Method & $^\sharp$Shots$^{[1]}$ & Acceleration$^{[1]}$ \\
\hline
$d=1$ & $\varepsilon=0.01$ & RSR & 1e6 & $\times$ 1\\
&  & ARSR & 1e5 & $\times$ 10\\
&  & FSR & 2.2e5 & $\times$ 5\\
\cline{2-5}
& $\varepsilon=0.001$ & RSR & 1e9 & $\times$ 1\\
&  & ARSR & 3.2e7 & $\times$ 31\\
&  & FSR & 1e8 & $\times$ 10\\
\hline
$d=2$ & $\varepsilon=0.01$ & RSR & 1e8 & $\times$ 1\\
&  & ARSR & 1e6 & $\times$ 100\\
&  & FSR & 4.6e5 & $\times$ 2.2e2\\
\cline{2-5}
& $\varepsilon=0.001$ & RSR & 1e12 & $\times$ 1\\
&  & ARSR & 1e9 & $\times$ 1000\\
&  & FSR & 2.6e8 & $\times$ 3.8e3\\
\hline
$d=3$ & $\varepsilon=0.01$ & RSR & 1e10 & $\times$ 1\\
&  & ARSR & 1e7 & $\times$ 1000\\
&  & FSR & 9.9e5 & $\times$ 1.0e4\\
\cline{2-5}
& $\varepsilon=0.001$ & RSR & 1e15 & $\times$ 1\\
&  & ARSR & 3.2e10 & $\times$ 3.1e4\\
&  & FSR & 6.9e8 & $\times$ 1.4e6\\
\hline
\end{tabular}
}
\begin{flushleft}
\footnotesize [1] These columns only indicate the rough numbers of shots and the acceleration rates because the pre-factors depend on the properties (e.g., function norms) of the underlying function. 
\end{flushleft}
\end{table}

In Table \ref{appF:tab1}, we consider the worst case of a discontinuous function for which the parameter $s$ is $2$. In this case, the ARSR method has large errors due to the discontinuity if $M<N$. Thus, we take $M=N$ and then the ARSR method is the same as the RSR method. Besides, the FSR method has a worse order than the RSR/ARSR method if $d\le 2$, although it is better if $d\ge 3$.  

In Table \ref{appF:tab2}, we consider the practical cases when the function is continuous and piecewise $W^{2,1}$ (i.e., up to second-order weak derivatives are integrable) so that the parameter $s$ is smaller than $2/3$. The ARSR and the FSR methods outperform the RSR method, and the accelerations are more significant for larger dimensions and smaller error bounds.

\end{document}